\documentclass[reprint,twocolumn,pre,showpacs,amsmath,amssymb,aps]{revtex4-1}

\usepackage{natbib}	
\usepackage{graphicx,color,rotating}
\usepackage[latin1]{inputenc}
\usepackage{textcomp}
\usepackage{dcolumn}
\usepackage{bm}     
\usepackage{upgreek}
\usepackage[latin1]{inputenc}

\usepackage{array}
\newcolumntype{L}[1]{>{\raggedright\let\newline\\\arraybackslash\hspace{0pt}}m{#1}}
\newcolumntype{C}[1]{>{\centering\let\newline\\\arraybackslash\hspace{0pt}}m{#1}}
\newcolumntype{R}[1]{>{\raggedleft\let\newline\\\arraybackslash\hspace{0pt}}m{#1}}
\newcommand\Tstrut{\rule{0pt}{3.0ex}}         

\usepackage[section]{placeins} 
\newcommand{\notop}{{{}_{}}}

\renewcommand{\vec}[1]{\bm{#1}}

\newcommand{\ee}{\mathrm{e}}
\newcommand{\ii}{\mathrm{i}}
\newcommand{\eiwt}{\mathrm{e}^{-\ii\omega t}}
\newcommand{\epiwt}{\mathrm{e}^{\ii\omega t}}
\newcommand{\eiIIwt}{\mathrm{e}^{-\ii 2\omega t}}
\newcommand{\dm}{\mathrm{d}}
\newcommand{\dd}{\mathrm{d}}
\newcommand{\avr}[1]{\big\langle #1 \big\rangle}
\newcommand{\IIw}[1]{\big( #1 \big)^{2\omega}}

\DeclareMathOperator{\re}{Re}

\newcommand{\tauE}{\tau^{{}}_E}
\newcommand{\taunu}{\tau^{{}}_\nu}

\newcommand{\pp}{\partial^{{}}}
\newcommand{\ppsqr}{\partial^{\,2_{}}}

\newcommand{\nablabf}{\boldsymbol{\nabla}}

\newcommand{\divop}{\nablabf\cdot}

\newcommand{\etal}{\textit{et~al.}}

\newcommand{\nnn}{\vec{n}}

\newcommand{\rrr}{\vec{r}}

\newcommand{\vvv}{\vec{v}}

\newcommand{\vstr}{v^{{}}_\mathrm{str}}

\newcommand{\zerovec}{\boldsymbol{0}}

\newcommand{\Eac}{E^{{}}_\mathrm{ac}}

\newcommand{\kaps}{\kappa^\notop_s}

\newcommand{\vbc}{v^\notop_\mathrm{bc}}
\newcommand{\vbcsqr}{v^2_\mathrm{bc}}

\newcommand{\etaB}{\eta^\mathrm{b}}
\newcommand{\etaBO}{\eta^\mathrm{b}_0}

\newcommand{\etab}{\eta^\mathrm{b}}

\newcommand{\etaO}{\eta^{{}}_0}

\newcommand{\gam}{\Gamma}

\newcommand{\omgO}{{\omega^\notop_0}}
\newcommand{\omgOsqr}{{\omega^2_0}}

\newcommand{\cs}{c^{{}}_s}
\newcommand{\cssqr}{c^{2}_s}

\newcommand{\fres}{f^{{}}_\mathrm{res}}

\newcommand{\gO}{g^{{}}_0}
\newcommand{\gI}{g^{{}}_1}
\newcommand{\gII}{g^{{}}_2}

\newcommand{\kO}{k^{{}}_0}
\newcommand{\kOsqr}{k^{2_{}}_0}

\newcommand{\pI}{p^{{}}_1}

\newcommand{\pIsqr}{p^{\,2_{}}_1}

\newcommand{\pII}{p^{{}}_2}
\newcommand{\pIIw}{p^{2\omega}_2}

\newcommand{\tO}{t^{{}}_0}

\newcommand{\tpw}{t^{{}}_\mathrm{pw}}

\newcommand{\taubf}{\bm{\tau}}
\newcommand{\taubfI}{\bm{\tau}^{{}}_1}
\newcommand{\taubfII}{\bm{\tau}^{{}}_2}

\newcommand{\vvvO}{\vvv^{{}}_0}

\newcommand{\vI}{v^{{}}_1}
\newcommand{\vIsqr}{v^{\,2_{}}_1}

\newcommand{\vvvI}{\vvv^{{}}_1}

\newcommand{\vyI}{v^{{}}_{y1}}
\newcommand{\vzI}{v^{{}}_{z1}}

\newcommand{\vII}{v^{{}}_2}

\newcommand{\vyII}{v^{{}}_{y2}}
\newcommand{\vzII}{v^{{}}_{z2}}
\newcommand{\vvvII}{\vvv^{{}}_2}
\newcommand{\vvvIIO}{\vvv^0_2}
\newcommand{\vvvIIw}{\vvv^{2\omega}_2}
\newcommand{\vvvIw}{\vvv^{\omega}_1}
\newcommand{\vIIO}{v^0_2}
\newcommand{\vIIw}{v^{2\omega}_2}
\newcommand{\vIw}{v^{\omega}_1}

\newcommand{\vIIwinhom}{v^{2\omega,\mathrm{inhom}}_2}

\newcommand{\rhoO}{\rho^\notop_0}
\newcommand{\rhoI}{\rho^\notop_1}

\newcommand{\SIm}{\textrm{m}}

\newcommand{\SImum}{\textrm{\textmu{}m}}

\newcommand{\SInm}{\textrm{nm}}

\newcommand{\SIs}{\textrm{s}}

\newcommand{\beq}[1]{\begin{equation} \eqlab{#1}}
\newcommand{\eeq}{\end{equation}}
\newcommand{\bsub}{\begin{subequations}}
\newcommand{\esub}{\end{subequations}}
\def\bal#1\eal{\begin{align}#1\end{align}}
\def\bsubal#1\esubal{\bsub \begin{align}#1\end{align} \esub}
\newcommand{\nn}{\nonumber}
\newcommand{\eqlab}[1]{\label{eq:#1}}
\renewcommand{\eqref}[1]{Eq.~(\ref{eq:#1})}
\newcommand{\eqsref}[2]{Eqs.~(\ref{eq:#1}) and~(\ref{eq:#2})}
\newcommand{\refeq}[1]{(\ref{eq:#1})}

\newcommand{\figref}[1]{Fig.~\ref{fig:#1}}

\newcommand{\figlab}[1]{\label{fig:#1}}

\newcommand{\secref}[1]{Section~\ref{sec:#1}}

\newcommand{\seclab}[1]{\label{sec:#1}}
\newcommand{\tabref}[1]{Table~\ref{tab:#1}}

\newcommand{\tablab}[1]{\label{tab:#1}}

\begin{document}

\title{A theoretical study of time-dependent, ultrasound-induced \\acoustic streaming in microchannels}

\author{Peter Barkholt Muller}
\email{peter.b.muller@gmail.com}
\affiliation{Department of Physics, Technical University of Denmark, DTU Physics Building 309, DK-2800 Kongens Lyngby, Denmark}

\author{Henrik Bruus}
\email{bruus@fysik.dtu.dk}
\affiliation{Department of Physics, Technical University of Denmark, DTU Physics Building 309, DK-2800 Kongens Lyngby, Denmark}

\date{Submitted to Phys.~Rev.~E, 8 September 2015}

\begin{abstract}
Based on first- and second-order perturbation theory, we present a numerical study of the temporal build-up and decay of unsteady acoustic fields and acoustic streaming flows actuated by vibrating walls in the transverse cross-sectional plane of a long straight microchannel under adiabatic conditions and assuming temperature-independent material parameters. The unsteady streaming flow is obtained by averaging the time-dependent velocity field over one oscillation period, and as time increases, it is shown to converge towards the well-known steady time-averaged solution calculated in the frequency domain. Scaling analysis reveals that the acoustic resonance builds up much faster than the acoustic streaming, implying that the radiation force may dominate over the drag force from streaming even for small particles. However, our numerical time-dependent analysis indicates that pulsed actuation does not reduce streaming significantly due to its slow decay. Our analysis also shows that for an acoustic resonance with a quality factor $Q$, the amplitude of the oscillating second-order velocity component is $Q$ times larger than the usual second-order steady time-averaged velocity component. Consequently, the well-known criterion $\vI\ll\cs$ for the validity of the perturbation expansion is replaced by the more restrictive criterion $\vI\ll\cs/Q$. Our numerical model is available in the supplemental material in the form of Comsol model files and Matlab scripts.
\end{abstract}


\maketitle


\section{Introduction}
\seclab{Intro}
Acoustophoresis has successfully been used in many applications to manipulate particles in the size range from about 0.5~mm down to about $2~\SImum$ \cite{Bruus2011c}. However, for smaller particles, the focusing by the acoustic radiation force is hindered by the drag force from the suspending liquid, which is set in motion by the generation of an acoustic streaming flow \cite{Muller2012, Barnkob2012a}. This limits the use of acoustophoresis to manipulate sub-micrometer particles, relevant for application within medical, environmental, and food sciences, and it underlines a need for better understanding of acoustic streaming and ways to circumvent this limitation.

The phenomenon of acoustic streaming was first described theoretically by Lord Rayleigh \cite{LordRayleigh1884} in 1884, and has later been revisited, among others, by Schlicting \cite{Schlichting1932}, Nyborg \cite{Nyborg1958}, Hamilton \cite{Hamilton2003, Hamilton2003a}, Rednikov and Sadhal \cite{Rednikov2011}, and Muller \etal\ \cite{Muller2013}, to extend the fundamental treatment of the governing equations and to solve the equations for various open and closed geometries.

Numerical methods have been applied in many studies to predict the streaming phenomena observed in various experiments. Muller \etal\ \cite{Muller2012} developed a numerical scheme to solve the acoustic streaming in the cross section of a long straight microchannel, which resolved the viscous acoustic boundary layers and described the interplay between the acoustic scattering force and the streaming-induced drag force on suspended particles. This scheme was later extended to take into account the thermoviscous effects arising from the dependence of the fluid viscosity on the oscillating temperature field \cite{Muller2014}.
Lei \etal \ \cite{Lei2013,Lei2014} have developed a numerical scheme based on the effective slip-velocity equations, originally proposed by Nyborg in 1953 \cite{Nyborg1953a, Lee1989}, which avoid the resolution of the thin boundary layers but still enable qualitative predictions of the three-dimensional streaming flows observed in microchannels and flat microfluidic chambers.
To obtain quantitative results from such models that does not resolve the acoustic boundary layers, Hahn \etal\ \cite{Hahn2015} developed an effective model to determine the loss associated with the viscous stresses inside the thermoacoustic boundary layers, and apply this loss as an artificial bulk absorption coefficient. This enables the calculation of correct acoustic amplitudes, without resolving the thin acoustic boundary layers.
Acoustic streaming in the cross section of a straight PDMS microchannel exited by surface acoustic waves was studied numerically by Nama \etal\ \cite{Nama2015}, describing the influence of the acoustically soft PDMS wall on the particle focusability, and examining the possibilities of having two tunable counter-propagating surface acoustic waves.

All of the above mentioned studies consider steady acoustic streaming flows. This is reasonable as the streaming flow reaches steady state typically in a few milliseconds, much faster than other relevant experimental timescales. Furthermore, this allows for analytical solutions for the streaming velocity field in some special cases, and it makes it much easier to obtain numerical solutions. However, an experimental study by Hoyos and Castro \cite{Hoyos2013} indicates that a pulsed actuation, instead of steady, can reduce the drag force from the streaming flow relative to the radiation force and thus allowing the latter also to dominate manipulation of sub-micrometer particles. This might provide an alternative method to the one proposed by Antfolk \etal \ \cite{Antfolk2014}, which used an almost square channel with overlapping resonances to create a streaming flow that did not counteract the focusing of sub-micrometer particles.

To theoretically study the effects of a pulsed ultrasound actuation, we need to solve the temporal evolution of the acoustic resonance and streaming, which is the topic of the present work. Numerical solutions of the time-domain acoustic equations were used by Wang and Dual \cite{Wang2009} to calculate the time-averaged radiation force on a cylinder and the steady streaming around a cylinder, both in a steady oscillating acoustic field. However, they did not present an analysis of the unsteady build-up of the acoustic resonance and the streaming flow.

In this paper, we derive the second-order perturbation expansion of the time-dependent governing equations for the acoustic fields and streaming velocity, and solve them numerically for a long straight channel with acoustically hard walls and a rectangular cross section. The analysis and results are divided into two sections: (1) A study of the transient build-up of the acoustic resonance and streaming from a initially quiescent state towards a steady oscillating acoustic field and a steady streaming flow. (2) An analysis of the response of the acoustic field and the streaming flow to pulsed actuation, and quantifying whether this can lead to better focusability of sub-micrometer particles.

In previous studies, such as \cite{Muller2012,Muller2014,Nama2015}, only the periodic state of the acoustic resonance and the steady time-averaged streaming velocity are solved. When solving the time-dependent equations, we obtain a transient solution, which may also be averaged over one oscillation period to obtain an unsteady time-averaged solution.

\section{Basic adiabatic acoustic theory}
\seclab{theory}
In this section we derive the governing equations for the first- and second-order perturbations to unsteady acoustic fields in a compressible Newtonian fluid. We only consider acoustic perturbation in fluids, and treat the surrounding solid material as ideal rigid walls. Our treatment is based on textbook adiabatic acoustics \cite{Pierce1991} and our previous study Ref. \cite{Muller2014} of the purely periodic state.


\begin{table}[t]
\caption{\tablab{parameter_values}
IAPWS parameter values for pure water at ambient temperature 25\:$^\circ$C and pressure $0.1013$~MPa. For references see Sec. II-B in Ref. \cite{Muller2014}.}
\begin{tabular}{ L{3.08cm} c  r@{ $\times$ }l  c }
\hline \hline
Parameter & Symbol & \multicolumn{2}{c}{Value} & Unit \\ \hline
\multicolumn{5}{l}{\textbf{Acoustic properties:}}  \Tstrut \\[0.5mm]
Mass density
 &
$\rhoO$
 &
$9.971$ & $10^{2}$
 &
kg\:m$^{-3}$
\\[1.0mm]
Speed of sound
 &
$c^{{}}_s$
 &
$1.497$ & $10^{3}$
 &
m\:s$^{-1}$
\\[1.0mm]
Compressibility
 &
$\kaps$ 
 &
$4.477$ & $10^{-10}$
 &
Pa$^{-1}$
\\[1.0mm]
\multicolumn{5}{l}{\textbf{Transport properties:}}   \\[0.5mm]
Shear viscosity
 &
$\eta$
 &
$8.900$ & $10^{-4}$
 &
Pa\:s
\\[1.0mm]
Bulk viscosity
 &
$\eta^\mathrm{b}$
 &
$2.485$ & $10^{-3}$
 &
Pa\:s
\\[2.5mm]
\hline \hline
\end{tabular}
\end{table}
%
%
\subsection{Adiabatic thermodynamics}
We employ the adiabatic approximation, which assumes that the entropy is conserved for any small fluid volume \cite{Landau1980}. Consequently, the thermodynamic state of the fluid is described by only one independent thermodynamic variable, which we choose to be the pressure $p$. See \tabref{parameter_values} for parameter values. The changes $\dd\rho$ in the density $\rho$ from the equilibrium state is given by
 \beq{stateEq}
 \dd \rho = \rho\kaps\, \dd p,
 \eeq
where the isentropic compressibility $\kaps$ is defined as
\beq{kaps}
\kaps = \frac{1}{\rho} \bigg(\frac{\pp \rho}{\pp p}\bigg)_s = \frac{1}{\rho\cssqr}.
\eeq
%

%
%
\subsection{Governing equations}
\seclab{GovEq}

Mass conservation implies that the rate of change $\pp_t\rho$ of the density in a test volume with surface normal vector $\nnn$ is given by the influx (direction $-\nnn$) of the mass current density $\rho\vvv$. In differential form by Gauss's theorem it is
 \bsub\eqlab{contEq}
 \beq{contEqa}
 \pp_t \rho = \nablabf\cdot\big[-\rho\vvv\big].
 \eeq
Substituting $\pp_t \rho$ and $\nablabf \rho$ using \eqref{stateEq}, and dividing by $\rho$, the continuity equation \refeq{contEqa} becomes
 \beq{contEqb}
 \kaps\pp_t p = -\divop\vvv - \kaps\vvv\cdot \nablabf p.
 \eeq
 \esub
Similarly, momentum conservation implies that the rate of change $\pp_t(\rho\vvv)$ of the momentum density in the same test volume is given by the stress forces $\bm{\sigma}$ acting on the surface (with normal $\nnn$), and the influx (direction $-\nnn$) of the momentum current density $\rho\vvv\vvv$. In differential form, neglecting body forces, this becomes
\bsub\eqlab{momentumEq}
 \beq{momentumEqa}
 \pp_t (\rho\vvv) = \nablabf\cdot\big[\bm{\tau} - p\:\bm{1} - \rho\vvv\vvv\big],
 \eeq
where the viscous stress tensor is defined as
 \beq{tauDef}
 \bm{\tau} = \eta\bigg[\nablabf\vvv + (\nablabf \vvv)^\mathrm{T}\bigg]
 + \bigg[\etaB -\frac{2}{3}\eta\bigg](\nablabf\cdot\vvv)\:\bm{1}.
 \eeq
Here $\bm{1}$ is the unit tensor and the superscript "T" indicates tensor transposition. Using the continuity equation \refeq{contEqa}, the momentum equation \refeq{momentumEqa} is rewritten into the well-known Navier--Stokes form,
 \beq{momentumEqb}
 \rho\pp_t \vvv = \nablabf\cdot\big[\bm{\tau} - p\:\bm{1}\big] - \rho(\vvv\cdot\nablabf)\vvv,
 \eeq \esub
which is useful when solving problems in the time domain. The equations \refeq{contEqb} and \refeq{momentumEqb} constitutes the non-linear governing equations which we will study by applying the usual perturbation approach of small acoustic amplitudes.

\subsection{First-order time-domain equations}
\seclab{FirstOrder}
The homogeneous, isotropic, quiescent thermodynamic equilibrium state is taken to be the zeroth-order state in the acoustic perturbation expansion. Following standard perturbation theory, all fields $g$ are written in the form $g = \gO + \gI$, for which $\gO$ is the value of the zeroth-order state, and $\gI$ is the acoustic perturbation which by definition has to be much smaller than $\gO$. For the velocity, the value of the zeroth-order state is $\vvvO = \zerovec$, and thus $\vvv = \vvvI$. The zeroth-order terms solve the governing equations in the zeroth-order state and thus drop out of the equations. Keeping only first-order terms, we obtain the following first-order equations.

The first-order continuity equation~(\ref{eq:contEqb}) becomes
 \beq{massEq1}
 \kaps\pp_t\pI = -\nablabf\cdot\vvvI,
 \eeq
and likewise, the momentum equation~(\ref{eq:momentumEqb})  becomes
 \bsub\eqlab{momentumEq1}
 \beq{momentumEq1part}
 \rhoO\pp_t\vvvI = \nablabf\cdot\big[\taubfI - \pI\bm{1}\big],
 \eeq
where $\taubfI$ is given by
 \beq{tau1}
\taubfI = \etaO\bigg[\nablabf\vvvI + (\nablabf \vvvI)^\mathrm{T}\bigg]
 + \bigg[\etaBO -\frac{2}{3}\etaO\bigg](\nablabf\cdot\vvvI)\:\bm{1}.
 \eeq
 \esub
Equations \refeq{massEq1} and \refeq{momentumEq1} determine together with a set of boundary conditions the time evolution of the first-order acoustic fields $\pI$ and $\vvvI$.
%
%
%
\subsection{Second-order time-domain equations}
\seclab{SecondOrder}

Moving on to second-order perturbation theory, we write the fields as $g = \gO + \gI + \gII$, with $\gI$ and $\gII$ depending on both time and space. For simplicity and in contrast to Ref. \cite{Muller2014}, we do not include perturbations in $\eta$ and $\etab$. This will cause the magnitude of the streaming to be slightly off, as does the adiabatic approximation, however the qualitative behavior is not expected to change. The second-order time-domain continuity equation~(\ref{eq:contEqb}) becomes
 \beq{massEq2}
 \kaps \pp_t \pII = - \divop\vvvII - \kaps\vvvI\cdot\nablabf \pI,
 \eeq
and the momentum equation~(\ref{eq:momentumEqb}) takes the form
 \bsub\eqlab{momentumEq2}
 \beq{momentumEq2A}
 \rhoO\pp_t\vvvII =  -\rhoI\pp_t\vvvI + \nablabf\cdot\big[\taubfII - \pII\:\bm{1} \big]- \rhoO(\vvvI\cdot\nablabf)\vvvI,
 \eeq
where $\taubfII$ is given by
\bal
 \eqlab{tau2}
 \taubfII &= \etaO\bigg[\nablabf\vvvII + (\nablabf \vvvII)^\mathrm{T}\bigg]
 + \bigg[\etaBO -\frac{2}{3}\etaO\bigg](\nablabf\cdot\vvvII)\:\bm{1} .
  \eal
Using \eqref{stateEq} in the form $\rhoI = \rhoO\kaps\pI$ and the first-order momentum equation \refeq{momentumEq1part}, we rewrite \eqref{momentumEq2A} to
 \bal\eqlab{momentumEq2B}
 \rhoO\pp_t\vvvII =  &\divop\big[\taubfII - \pII\:\bm{1} - \kaps\pI\taubfI + \tfrac{1}{2}\kaps\pIsqr\:\bm{1} \big]\nonumber\\
 &+\kaps\nablabf\pI\cdot\taubfI - \rhoO(\vvvI\cdot\nablabf)\vvvI.
 \eal
 \esub
This particular form of the second-order momentum equation is chosen to minimize numerical errors as described in \secref{numgov}.

%
\subsection{Periodic frequency-domain equations}
\seclab{freqdom}
When solving for the periodic state at ${t\rightarrow \infty}$, it is advantageous to formulate the first-order equations in the frequency domain. The harmonic first-order fields are all written as  ${g^{{}}_1(\rrr,t) = \re\big\{g_1^\mathrm{fd}(\rrr)\ee^{-\ii \omega t}\big\}}$, where $g_1^\mathrm{fd}$ is the complex field amplitude in the frequency domain.
The first-order frequency-domain equations are derived from \eqsref{massEq1}{momentumEq1part} by the substitution $\pp_t \rightarrow -\ii\omega$,
\bal
 \nablabf\cdot\vvv_1^\mathrm{fd} - \ii\omega\kaps p_1^\mathrm{fd} &= 0,\eqlab{freqdom1stmass}\\
 \nablabf\cdot\big[\taubf_1^\mathrm{fd} - p_1^\mathrm{fd}\bm{1}\big] + \ii\omega\rhoO\vvv_1^\mathrm{fd} &= \zerovec.\eqlab{freqdom1stmom}
\eal
The steady time-averaged streaming flow is obtained from the time-averaged second-order frequency-domain equations, where $\avr{g_2^\mathrm{fd}}$ denotes time averaging over one oscillation period of the periodic second-order field.
The time-average of products of two harmonic first-order fields $g_1^\mathrm{fd}$ and $\tilde{g}_1^\mathrm{fd}$ is given by ${\avr{g_1^\mathrm{fd} \tilde{g}_1^\mathrm{fd}} = \frac{1}{2}\re\big[\big(g_1^\mathrm{fd}\big)^*\tilde{g}_1^\mathrm{fd}\big]}$, as in Ref. \cite{Muller2014}, where the asterisk denotes complex conjugation.
In the periodic state, the fields may consist of harmonic terms and a steady term, and thus all full time-derivatives average to zero $\avr{\pp_t g_2^\mathrm{fd}} = 0$.
The time-averaged second-order frequency-domain equations are derived from \eqsref{massEq2}{momentumEqa},
\bal
 \divop\avr{\vvv_2^\mathrm{fd}} + \kaps\avr{\vvv_1^\mathrm{fd}\cdot \nablabf p_1^\mathrm{fd}} &= 0, \eqlab{freqdom2ndmass}\\
 \nablabf\cdot\big[\avr{\taubf_2^\mathrm{fd}} - \avr{p_2^\mathrm{fd}}\:\bm{1} - \rhoO\avr{\vvv_1^\mathrm{fd}\vvv_1^\mathrm{fd}}\big] &= \zerovec.\eqlab{freqdom2ndmom}
\eal
%

%
%
\subsection{Acoustic energy and cavity Q-factor}
\seclab{EnergyandQ}

The total acoustic energy of the system in the time domain $\Eac(t)$ and in the frequency domain $\avr{E_\mathrm{ac}^\mathrm{fd}(\infty)}$ is given by
\bsub\eqlab{Eacboth}
\bal\eqlab{Eac}
\Eac(t) &= \int_V \bigg[\frac{1}{2}\kaps \pIsqr + \frac{1}{2}\rhoO \vIsqr\bigg]\: \dd V,\\[2mm]
\avr{E_\mathrm{ac}^\mathrm{fd}(\infty)} &= \int_V \bigg[\frac{1}{2}\kaps \avr{p_1^\mathrm{fd} p_1^\mathrm{fd}}
+ \frac{1}{2}\rhoO \avr{\vvv_1^\mathrm{fd}\cdot \vvv_1^\mathrm{fd}}\bigg]\: \dd V \eqlab{Eacfd}.
\eal
\esub
Moreover, the time derivative of $\Eac(t)$ is
\bsub
\bal
\pp_t\Eac &= \int_V \pp_t\bigg[\frac{1}{2}\kaps \pIsqr + \frac{1}{2}\rhoO \vIsqr\bigg] \ \dd V \nonumber\\
& = \int_V \big[\kaps \pI \pp_t\pI + \rhoO \vvvI\cdot\pp_t\vvvI\big] \ \dd V\nonumber\\
&= \int_V \!\bigg\{\!\nablabf\cdot\Big[\vvvI\cdot(\taubfI\!-\!\pI\bm{1})\Big] - \nablabf\vvvI\! :\! \taubfI \bigg\} \dd V,\eqlab{dtEacB}
\eal
where we have used \eqsref{massEq1}{momentumEq1part}.
Applying Gauss's theorem on the first term in \eqref{dtEacB}, we arrive at
\bal
\pp_t\Eac &= \int_A \Big[\vvvI\cdot(\taubfI-\pI\bm{1})\Big]\cdot \nnn\ \dd A - \int_V \nablabf\vvvI : \taubfI \ \dd V \nonumber\\
&= P_\mathrm{pump} - P_\mathrm{dis}\eqlab{Ebalance},
\eal
\esub
where $P_\mathrm{pump}$ is the total power delivered by the forced vibration of the sidewalls, and $P_\mathrm{dis}$ is the total power dissipated due to viscous stress.
The quality factor $Q$ of a resonant cavity is given by
\bal
\eqlab{Qdef}
Q &= 2\pi\frac{\text{Energy stored}}{\text{Energy dissipated per cycle}} = \omega\frac{\avr{E_\mathrm{ac}^\mathrm{fd}}}{\avr{P_\mathrm{dis}^\mathrm{fd}}}.
\eal

%
\subsection{Summary of theory}
\seclab{Summary}
Throughout this paper we refer to two kinds of solutions of the acoustic energy and velocity fields: unsteady non-periodic solutions obtained from Eqs.~\refeq{massEq1}-\refeq{momentumEq2} and steady periodic solutions obtained from Eqs.~\refeq{freqdom1stmass}-\refeq{freqdom2ndmom}. When presenting the unsteady non-periodic solutions, they are often normalized by the steady periodic solution, to emphasize how close it has converged towards this solution.

\section{Numerical model}
The numerical scheme solves the governing equations for the acoustic field inside a water domain enclosed by a two-dimensional rectangular microchannel cross section. The vibrations in the surrounding chip material and piezo transducer are not modeled. The water domain is surrounded by immovable hard walls, and the acoustic field is excited by oscillating velocity boundary conditions, representing an oscillating nm-sized displacement of the walls. A sketch of the numerical model is shown in \figref{chipmesh}(a). We exploit the symmetry along the horizontal center axis $z=0$, reducing our computational domain by a factor of two. The system is also symmetric about the vertical center axis $y=0$, however, our attempts to use this symmetry introduced numerical errors, and consequently it was not exploited in the numerical model. The model used to calculate the steady streaming flow in the time-periodic case is a simplification of the model presented in Ref. \cite{Muller2014}, whereas the model used to solve the time-dependent problem is new.

\begin{figure}
\centering
\includegraphics[width=0.99\columnwidth]{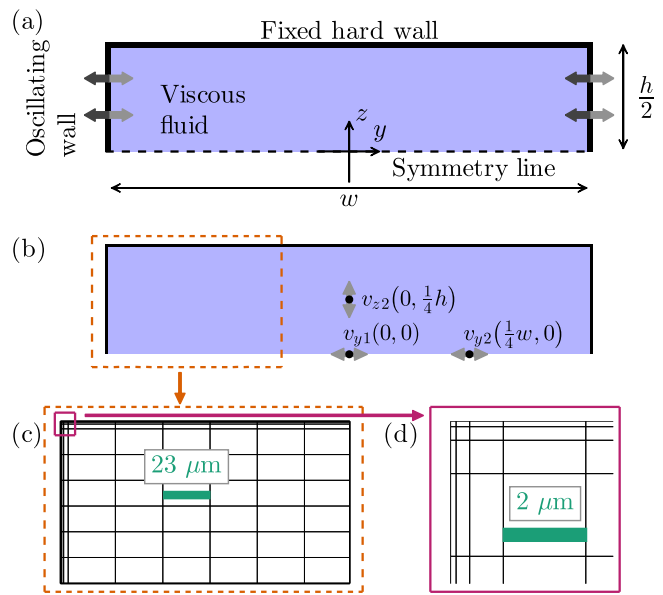}
\caption{\figlab{chipmesh} (Color online) (a) Sketch of the rectangular computational domain in the $yz$-plane representing the upper half of a rectangular cross section of a long straight microchannel of width $w=380\ \SImum$ and height $h=160\ \SImum$ as in \cite{Augustsson2011}.  The thick arrows indicate in-phase oscillating velocity actuation at the left and right boundaries. (b) The three black points indicate positions at which the velocity components (gray arrows), defined in \eqref{decomtime}, are probed. (c) Sketch of the spatial mesh used for the discretization of the physical fields. (d) A zoom-in on the mesh in the upper left corner.}
\end{figure}

\subsection{Governing equations}
\seclab{numgov}
The governing equations are solved using the commercial software Comsol Multiphysics \cite{COMSOL44} based on the finite element method \cite{Brenner2008}. To achieve greater flexibility and control, the equations are implemented through mathematics-weak-form-PDE modules and not through the built-in modules for acoustics and fluid mechanics. The governing equations are formulated to avoid evaluation of second-order spatial derivatives and of time-derivatives of first-order fields in the second-order equations, as time-derivatives carry larger numerical errors compared to the spatial derivatives. To fix the numerical solution of the second-order equations, a zero spatial average of the second-order pressure is enforced by a Lagrange multiplier. For the time-domain simulations we use the generalized alpha solver \cite{comsolref}, setting the alpha parameter to 0.5 and using a fixed time step $\Delta t$. Furthermore, to limit the amount of data stored in Comsol, the simulations are run from Matlab \cite{Matlab2012a} and long time-marching schemes are solved in shorter sections by Comsol.
Comsol model files and Matlab scripts are provided in the Supplemental Material \footnote{See Supplemental Material at [URL] for Comsol model files, both in a simple version \texttt{Muller2015\_TimeDepAcoust\_simple.mph}, and a full version \texttt{Muller2015\_TimeDepAcoust\_full.mph}, that allows for sectioning in smaller time intervals using the supplied Matlab script \texttt{Muller2015\_TimeDepAcoust\_full.m}}.

\subsection{Boundary conditions}
\seclab{BC}
The acoustic cavity is modeled with stationary hard rigid walls, and the acoustic fields are exited on the side walls by an oscillating velocity boundary condition with oscillation period~$\tO$ and angular frequency $\omega$,
 \beq{t0_def}
 \tO = \frac{2\pi}{\omega}.
 \eeq
The symmetry of the bottom boundary is described by zero orthogonal velocity component and zero orthogonal gradient of the parallel velocity component. The explicit boundary conditions for the first-order velocity become
\bsub\eqlab{BCv1}
\bal
&\text{top:}  &\vyI&=0,\quad &\vzI&=0,\\
&\text{bottom:}  & \pp_z v^\notop_{y1}&=0,\quad &v^\notop_{z1}&=0, \\
&\text{left-right:}  & v^\notop_{y1}&=\vbc\sin(\omega t),\quad &v^\notop_{z1}&=0.
\eal
\esub
The boundary conditions on the second-order velocity are set by the zero-mass-flux condition $\nnn \cdot \rho\vvv = 0$ on all boundaries, as well as zero parallel velocity component on the top, right and left wall boundaries, and zero orthogonal derivative of the parallel component of the mass flux on the bottom symmetry boundary. The explicit boundary conditions for the second-order velocity become
\bsub\eqlab{BCv2}
\bal
&\text{top:} &\vyII&=0,\quad &\vzII&=0,\\
&\text{bottom:} \hspace{3mm} & \pp_z \big(\rhoO v^\notop_{y2} + \rhoI v^\notop_{y1}\big)&=0,\quad &v^\notop_{z2}&=0, \\
&\text{left-right:} & \rhoO v^\notop_{y2} + \rhoI v^\notop_{y1}&=0,\quad &v^\notop_{z2}&=0.
\eal
\esub
%

\subsection{Spatial resolution}
The physical fields are discretized using fourth-order basis functions for $\vvvI$ and $\vvvII$ and third-order basis functions for $\pI$ and $\pII$. The domain shown in \figref{chipmesh}(a) is covered by basis functions localized in each element of the spatial mesh shown in \figref{chipmesh}(c). Since the streaming flow is solved in the time domain, the computational time quickly becomes very long compared to the computational time of solving the usual steady streaming flow. Thus we have optimized the use of precious few mesh elements to obtain the best accuracy of the solution. We use an inhomogeneous mesh of rectangular elements ranging in size from 0.16~$\SImum$ at the boundaries to 24~$\SImum$ in the bulk of the domain.
The convergence of the solution $g$ with respect to a reference solution $g_\mathrm{ref}$ was considered through the relative convergence parameter $C(g)$ defined in Ref.~\cite{Muller2014} by
\beq{C}
 C(g) = \sqrt{\frac{{\displaystyle \int} \big(g-g_\mathrm{ref}\big)^2\ \dm y\:\dm z}{
 {\displaystyle \int} \big(g_\mathrm{ref}\big)^2\ \dm y\:\dm z}}.
\eeq
In Ref.~\cite{Muller2014}, $C(g)$ was required to be below 0.001 for the solution to have converged.
The solution for the steady time-averaged velocity $\avr{\vvv_2^\mathrm{fd}(\infty)}$, calculated with the mesh shown in \figref{chipmesh}(c) and \ref{fig:chipmesh}(d), has $C = 0.006$ with respect to the solution calculated with the fine triangular reference mesh in Ref. \cite{Muller2014}, which is acceptable for the present study.

%
\subsection{Temporal resolution}

The required temporal resolution for time-marching schemes is normally determined by the Courant--Friedrichs--Lewy (CFL) condition \cite{comsolnonlinearacoustics}, also referred to as just the Courant number
\beq{CFL}
\mathrm{CFL} = \frac{\cs\,\Delta t}{\Delta r} \leq \mathrm{CFL}_\mathrm{max},
\eeq
where $\Delta t$ is the temporal discretization and $\Delta r$ is the spatial discretization. This means that the length over which a disturbance travels within a time step $\Delta t$ should be some fraction of the mesh element size, ultimately ensuring that disturbances do not travel through a mesh element in one time step. A more accurate interpretation of the CFL-condition is that it ensures that the error on the approximation of the time-derivative is smaller than the error on the approximation of the spatial-derivatives.
Consequently, the value of $\mathrm{CFL}_\mathrm{max}$ depends on the specific solver and on the order of the basis functions. For fourth-order basis functions and the generalized alpha solver, Ref.~\cite{comsolnonlinearacoustics} reports a value of $\mathrm{CFL}_\mathrm{max}^\mathrm{4th}=0.05$, which is an empirical result for a specific model.
Due to the inhomogeneity of the mesh, two values for the upper limit for the temporal resolution can be calculated based on \eqref{CFL}; $\Delta t = 8\times10^{-10}\ \mathrm{ns} \approx \tO/600$ for the bulk mesh size of 24~\SImum\ and $\Delta t = 5\times10^{-12}\ \mathrm{ns} \approx \tO/95000  $ for the boundary mesh size of 160~nm.

\begin{figure}
\centering
\includegraphics[width=0.99\columnwidth]{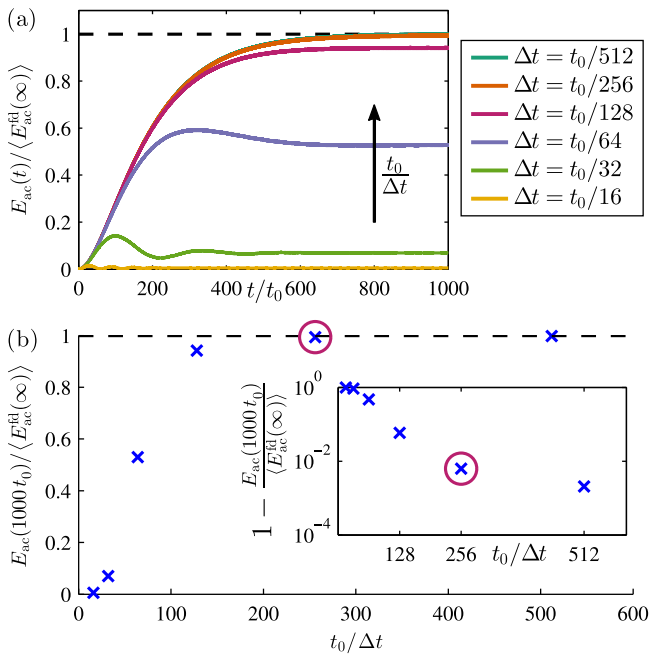}
\caption{\figlab{temporal} (Color online) Numerical convergence and temporal resolution. (a) Graphs of the build-up of acoustic energy $\Eac(t)$ in the time-domain simulations calculated with different fixed time steps $\Delta t$. The energy of the time-domain simulations is normalized with respect to the energy $\avr{E_\mathrm{ac}^\mathrm{fd}(\infty)}$ of the steady solution in the frequency domain, and should thus converge towards unity. In all simulations the actuation frequency equals the resonance frequency discussed in \secref{resonance}. (b) Acoustic energy $E_\mathrm{ac}(1000\,\tO)$ at $t=1000\,\tO$, normalized by $\avr{E_\mathrm{ac}^\mathrm{fd}(\infty)}$, and plotted versus the temporal resolution $\tO/\Delta t$ of the oscillation. The inset is a semilog plot of the relative deviation of $E_\mathrm{ac}(1000\,\tO)$ from $\avr{E_\mathrm{ac}^\mathrm{fd}(\infty)}$. The circled point in each graph indicates the time step $\Delta t = \tO/256$ used in all subsequent simulations.}
\end{figure}

To determine a reasonable trade-off between numerical accuracy and computational time, we study the convergence of the transient solution towards the steady solution for different values of the temporal resolution $\tO/\Delta t$. The acoustic energy $\Eac(t)$ is shown in \figref{temporal}(a) for different values of $\Delta t$ and normalized by the steady time-averaged energy $\avr{E_\mathrm{ac}^\mathrm{fd}(\infty)}$ of the frequency-domain calculation, and it is thus expected to converge to the unity for long times. In \figref{temporal}(b), $\Eac(1000\tO)/\avr{E_\mathrm{ac}^\mathrm{fd}(\infty)}$ is plotted versus the temporal resolution $\tO/\Delta t$, which shows how the accuracy of the time-domain solution increases as the temporal resolution is increased. In all subsequent simulations we have chosen a time step of $\Delta t = \tO/256$, the circled point in \figref{temporal}(b), for which the time-domain energy converge to 99.4\% of the energy of the steady calculation. The chosen value for the time step is larger than the upper estimate $\tO/600$ of the necessary $\Delta t$ based on the CFL-condition. This might be because our spatial domain is smaller than the wavelength, and consequently a finer spatial resolution is needed, compared to what is usually expected to spatially resolve a wave.

We have noted that the fastest convergence is obtained when actuating the system at its (numerically determined) resonance frequency $\fres$. When shifting the actuation frequency half the resonance width $\frac12\Delta f$ away from $\fres$, the energy $\Eac(t)$ for ${\Delta t = \tO/256}$ converged to only 95\% of the steady value $\avr{E_\mathrm{ac}^\mathrm{fd}(\infty)}$ (calculated in the frequency domain), thus necessitating smaller time steps to obtain reasonable convergence.

The computations where performed on a desktop PC with Intel Xeon CPU X5690 3.47~GHz 2 processors, 64-bit Windows~7, and 128~GB RAM. The computations took approximately one hour for each time interval of width $100 \tO$ with $\Delta t=\tO/256$, and the computational time was not limited by RAM, as only less than 2 ~GB RAM was allocated by Comsol for the calculations.

\section{Onset of acoustic streaming}

In this section the fluid is initially quiescent. Then, at time $t=0$, the oscillatory velocity actuation is turned on, such that within the first oscillation period its amplitude increases smoothly from zero to its maximum value $\vbc$, which it maintains for the rest of the simulation. We study the resulting build-up of the acoustic resonance and the acoustic streaming flow.

\subsection{Resonance and build-up of acoustic energy}
\seclab{resonance}

To determine the resonance frequency, the steady acoustic energy  $\avr{E_\mathrm{ac}^\mathrm{fd}(\infty)}$ \eqref{Eacfd} was calculated for a range of frequencies based on the frequency-domain equations \refeq{freqdom1stmass}-\refeq{freqdom1stmom}. In \figref{resonance} the numerical results (circles) are shown together with a Gaussian fit (full line), while the inset exhibits the fitted resonance frequency $\fres$, the full width $\Delta f$ at half maximum, and the quality factor ${Q=\fres/\Delta f}$.

\begin{figure}[t]
\centering
\includegraphics[width=0.99\columnwidth]{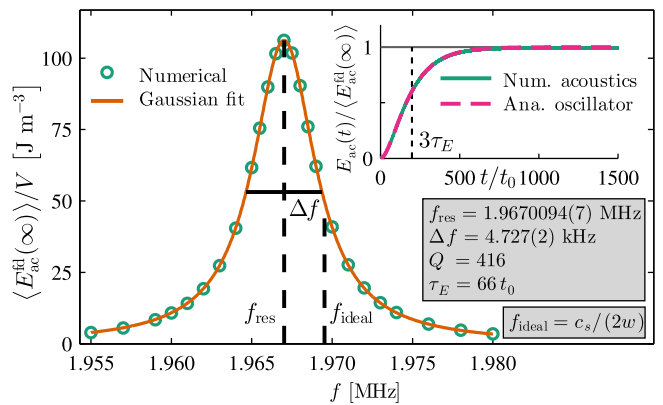}
\caption{\figlab{resonance} (Color online) Resonance curve and build-up of acoustic energy. (a) The numerical acoustic energy density ${\avr{E_\mathrm{ac}^\mathrm{fd}(\infty)}/V}$ (circles) for different frequencies of the boundary actuation and a Gaussian fit (full line) to the numerical data. $\fres$ is the fitted resonance frequency at the center of the peak, while $f_\mathrm{ideal}$ is the frequency corresponding to matching a half-wavelength with the channel width. The inset shows the numerical build-up of the acoustic energy (full line) for actuation at the resonance frequency, $\omega = 2\pi\fres$, along with the analytical prediction \eqref{oscillatorE} (dashed line) for a single harmonic oscillator with the same resonance frequency and quality factor $Q=\fres/\Delta f$.}
\end{figure}

The build-up of the acoustic energy in the cavity is well captured by a simple analytical model of a single sinusoidally-driven damped harmonic oscillator with time-dependent position $x(t)$,
\beq{oscillatorEq}
\frac{\mathrm{d}^2x}{\mathrm{d}t^2} + 2\gam\omgO\frac{\mathrm{d}x}{\mathrm{d}t} + \omgOsqr x = \frac{1}{m} F_0 \sin(\omega t).
\eeq
Here, $\gam$ is the non-dimensional loss factor, $\omgO$ is the resonance frequency of the oscillator, $\frac{1}{m}F_0$ is the amplitude of the driving force divided by the oscillator mass, and $\omega$ is the frequency of the forcing. The loss factor is related to the quality factor by $\gam=1/(2Q)$, and in the underdamped case $\gam<1$, the solution becomes
\bal
x(t) = A
\bigg[ &\sin(\omega t + \phi) \nonumber\\
&
- \frac{\omega\,\ee^{-\gam \omgO t}}{\omgO\sqrt{1-\gam^2}} \: \sin\left(\sqrt{1-\gam^2}\, \omgO t + \phi\right)\bigg]. \eqlab{oscillatorSol}
\eal
The amplitude $A$ and the phase shift $\phi$ between the forcing and the response are known functions of $\frac{F_0}{m}$, $\omgO$, $\omega$, and $\gam$, which are not relevant for the present study. From \eqref{oscillatorSol} we obtain the velocity $\dd x/\dd t$, leading to the total energy $E$ of the oscillator,
 \beq{oscillatorE}
 E = \tfrac{1}{2}m\omgOsqr x^2 + \tfrac{1}{2}m\bigg(\frac{\dd x}{\dd t}\bigg)^2.
 \eeq
Based on \eqsref{oscillatorSol}{oscillatorE}, the characteristic timescale $\tau^{{}}_E$ for the build-up of the acoustic energy is found to be
 \beq{tauE}
 \tau^{{}}_E = \frac{1}{2\gam\omgO} = \frac{Q}{\omgO}.
 \eeq
The build-up of the energy in the single harmonic oscillator, calculated at $\omega = \omgO$ with $\gam=1.20\times10^{-3}$, is shown in the inset of \figref{resonance} together with the build-up of acoustic energy $\Eac(t)$ of the microfluidic channel solved numerically at resonance, $\omega = 2\pi \fres$. The analytical and numerical results are in good agreement, and we conclude that the build-up of acoustic energy in the channel cavity can be modeled as a single harmonic oscillator. The energy builds up to 95\% of its steady value in about $500\,\tO\approx 8\,\tauE$.

%
\subsection{Decomposition of the velocity field}
The task of calculating the build-up of the acoustic streaming flow is a multi-scale problem, because the amplitude of the oscillating acoustic velocity field is several orders of magnitude larger than the magnitude of the streaming flow. This is indeed the very reason that we can apply the perturbation expansion
\bal
\vvv = \vvvI + \vvvII,
\eal
and decompose the non-linear governing equations into a set of linear first-order equations and a set of second-order equations. However, there is also another level of difference in velocity scaling. In the purely periodic state, the velocity can be Fourier decomposed as
\beq{decomfreq}
\vvv(\rrr,t) = \vvvIw(\rrr)\sin(\omega t) + \vvvIIw(\rrr)\sin(2\omega t) + \vvvIIO(\rrr),
\eeq
where $\vvvIw(\rrr)$ is the steady amplitude of the first-order harmonic component, $\vvvIIw(\rrr)$ is the steady amplitude of the second-order frequency-doubled component, and $\vvvIIO(\rrr)$ is the magnitude of the second-order steady velocity component referred to as the acoustic streaming flow. The orders of magnitude of the three velocity components in the periodic state are given by
\beq{decomAmp}
\vIw \sim Q\vbc,\quad \vIIw \sim \frac{Q^3\vbcsqr}{\cs},\quad \vIIO \sim \frac{Q^2\vbcsqr}{\cs}.
\eeq
The order of $\vI$ is derived in the one-dimensional acoustic cavity example presented in Ref. \cite{Bruus2012}, the order of $\vIIO$ is given by the well-known Rayleigh theory, while the order of $\vIIw$ is a new result derived in Appendix A.  The magnitude of $\vIIw$ is a factor of $Q$ larger than what is expected from dimensional analysis of the second-order equation \refeq{momentumEq2B}. Consequently, the criterion $|\vII|\ll|\vI|$ for the perturbation expansion becomes
\beq{perturbation2nd}
Q^2 \vbc\ll \cs,
\eeq
which is more restrictive than the usual criterion based on the first-order perturbation expansion, $Q\vbc\ll\cs$. Thus, the perturbation expansion becomes invalid for smaller values of $\vbc$ than previously expected.

In the transient regime we cannot Fourier decompose the velocity field. Instead, we propose a decomposition using envelope functions inspired by \eqref{decomfreq},
\beq{decomtime}
\vvv(\rrr,t) = \vvvIw(\rrr,t)\sin(\omega t) + \vvvIIw(\rrr,t)\sin(2\omega t) + \vvvIIO(\rrr,t).
\eeq
Here, the amplitudes are slowly varying in time compared to the fast oscillation period $\tO$. We can no longer separate $\vvvIIw$ and $\vvvIIO$ before solving the second-order time-dependent equations \refeq{massEq2} and \refeq{momentumEq2}. To obtain the time-dependent magnitude of the quasi-steady streaming velocity mode $\vvvIIO$, we need to choose a good velocity probe, and we thus form the unsteady time-average of $\vvvII(\rrr,t)$,
\beq{timeint}
\avr{\vvvII(\rrr,t)} = \int_{t-\tO/2}^{t+\tO/2}\: \vvvII(\rrr,t')\: \dd t'.
\eeq
The time-averaging is done with a fifth-order Romberg integration scheme \cite{Press2002} using data points with a uniform spacing of $\tO/16$ in the time interval of width $\tO$.

%
%
\subsection{Steady and unsteady streaming flow}
In this section we compare the unsteady time-averaged second-order velocity field $\avr{\vvvII(\rrr,t)}$, from the time-domain simulations, with the steady time-averaged second-order velocity field $\avr{\vvv_2^\mathrm{fd}(\rrr,\infty)}$, from the frequency-domain simulation.
Figure \ref{fig:2Dstreaming}(a) and (b) each shows a snapshot in time of the transient $\vvvI$ and $\vvvII$, respectively. For $\vvvII(\rrr,t)$,  the oscillatory component $\vvvIIw(\rrr,t)\sin(2\omega t)$ dominates, as it is two orders of magnitude  larger than the quasi-steady component $\vvvIIO(\rrr,t)$. However, at late times, here ${t=3000\,\tO}$, the amplitude $\vvvIIw(\rrr,t)$ has converged, and in $\avr{\vvvII(\rrr,t)}$ the oscillatory component average to zero and only the quasi-steady component remains.

The unsteady time average $\avr{\vvvII(\rrr,t)}$ evaluated at ${t=3000\,\tO}$ is shown in \figref{2Dstreaming}(c), exhibiting a single flow roll, in agreement with the classical Rayleigh streaming flow. In \figref{2Dstreaming}(d) is shown the steady  $\avr{\vvv_2^\mathrm{fd}(\infty)}$ from the frequency-domain simulation. Figure \ref{fig:2Dstreaming}(c) and \ref{fig:2Dstreaming}(d) use the same color scaling for the velocity magnitude, to evaluate the convergence of the unsteady streaming flow $\avr{\vvvII(3000\,\tO)}$ towards the steady streaming flow  $\avr{\vvv_2^\mathrm{fd}(\infty)}$, and the two solutions agree well both qualitatively and quantitatively.
The convergence parameter $C$, \eqref{C}, of $\avr{\vvvII(3000\,\tO)}$ with respect to $\avr{\vvv_2^\mathrm{fd}(\infty)}$ is ${C=0.01}$, and if we multiply $\avr{\vvvII}$ by a free factor, taking into account that $\avr{\vvvII}$ has not fully converged at ${t=3000\,\tO}$, the convergence parameter can be reduced to ${C=0.008}$. The remaining small difference between the unsteady $\avr{\vvvII(3000\,\tO)}$ and the steady $\avr{\vvv_2^\mathrm{fd}(\infty)}$ is attributed to the finite temporal resolution of the time marching scheme. We can thus conclude that the time-domain streaming simulation converges well towards the frequency-domain simulation, and this constitutes the primary validation of the unsteady non-periodic simulations.

\begin{figure}[!b]
\centering
\includegraphics[width=0.99\columnwidth]{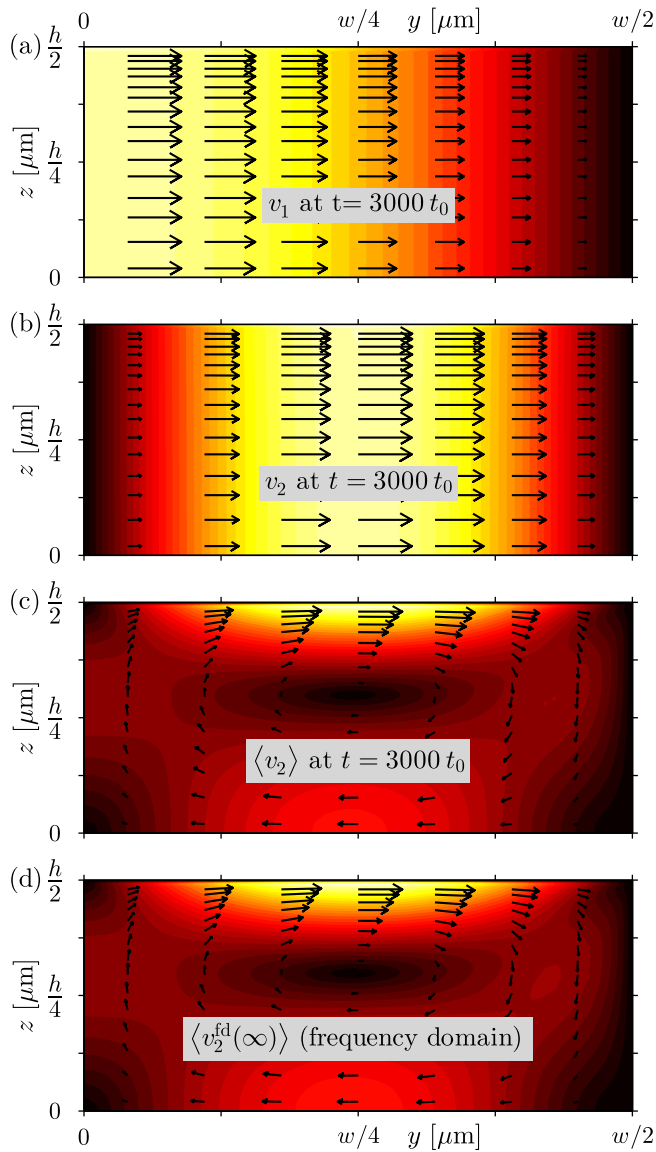}
\caption{\figlab{2Dstreaming} (Color online)
(a) Snapshot of the oscillatory first-order velocity field $\vvvI$ (vectors) and its magnitude [color plot ranging from 0 m/s (black) to 0.7 m/s (white)] at ${t=3000\,\tO}$. (b) Snapshot of the oscillatory second-order velocity field $\vvvII$ (vectors) and its magnitude [color plot ranging from 0 m/s (black) to 0.02 m/s (white)] at ${t=3000\,\tO}$. (c) Snapshot of the unsteady time-averaged second-order velocity field $\avr{\vvvII}$ (vectors), \eqref{timeint}, and its magnitude [color plot ranging from 0 mm/s (black) to 0.1 mm/s (white)] at ${t=3000\,\tO}$. (d) Steady time-averaged second-order velocity field $\avr{\vvv_2^\mathrm{fd}(\infty)}$ (vectors), \eqsref{freqdom2ndmass}{freqdom2ndmom}, and its magnitude [color scaling as in (c)]. In both the time-domain and the frequency-domain simulations the parameters of the oscillating velocity boundary condition was $\omega = 2\pi\fres$ and $\vbc = \omega d$, with wall displacement $d = 1\, \SInm$.}
\end{figure}

%
%
\subsection{Build-up of the velocity field}
To visualize the build-up of the acoustic fields over short and long timescales, we have chosen the three point probes shown in \figref{chipmesh}(b). The oscillating first-order velocity field is probed in the center of the channel $(0,0)$, far from the walls in order to measure the bulk amplitude of the acoustic field. The horizontal component of the second-order velocity $\vyII$ is probed on the horizontal symmetry axis at $(\frac{1}{4}w,0)$, where the oscillatory component $\vvvIIw$ has it maximum amplitude. The vertical component of the second-order velocity $\vzII$ is probed on the vertical symmetry axis at $(0,\frac{1}{4}h)$ where the oscillatory component $\vvvIIw$ is  small and of the same order as the quasi-steady component $\vvvIIO$, making the unsteady time-averaged second-order velocity at this point a good probe for the quasi-steady streaming velocity.

In \figref{short} is shown the build-up of the velocity probes (a-c) and their time-averages (d-f) for the first 20 oscillations. The thick lines are the oscillating velocities while, the thin lines are the envelopes of the oscillations. Already within the first 20 oscillation periods we see in \figref{short}(f) the build-up of a quasi-steady velocity component. The unsteady time-averaged horizontal velocity $\avr{\vyII}$, \figref{short}(e), is still primarily oscillatory, showing that for this probe the oscillatory component $\vvvIIw$ is much larger than the quasi-steady component $\vvvIIO$.

The temporal evolution of the velocity probes on the longer time scale up to $t=1500\,\tO$ is shown in \figref{long}. In \figref{long}(a) and (b) the amplitudes of the oscillatory first- and second-order velocity components are seen to stabilize around ${t =700\,\tO \sim 10\,\tauE}$. The steady amplitudes of the velocity probes \figref{long} agree with the theoretical predictions of \eqref{decomAmp}, yielding orders of magnitude $\vIw/\cs \sim 3\times10^{-4}$ (\figref{long}(a)), ${\vIIw/\cs \sim 5\times10^{-5}}$ (\figref{long}(b)), and ${\vIIO/\cs \sim 1\times10^{-7}}$ (\figref{long}(e) and \ref{fig:long}(f)).
The time-average of $\vyI$ tends to zero for long times as it is purely oscillatory, whereas the time-average of $\vyII$ tends to the magnitude of the quasi-steady component $\vvvIIO$, because the large but now steady oscillatory component $\vvvIIw$ average to zero. The dashed lines in \figref{long}(e) and (f) represent the magnitude of the steady time-averaged second-order velocity $\avr{\vvv_2^\mathrm{fd}(\infty)}$ from the frequency-domain simulation.

%
\section{Acoustic streaming generated by pulsed actuation}
\seclab{pulsed}
In the following we study the effects of switching the oscillatory boundary actuation on and off on a timescale much longer than the oscillation period $\tO$ in either single- or multi-pulse mode. The aim is to investigate whether such an approach can suppress the influence of the streaming flow on suspended particles relative to that of the radiation force.

\begin{figure*}
\centering
\includegraphics[width=0.99\textwidth]{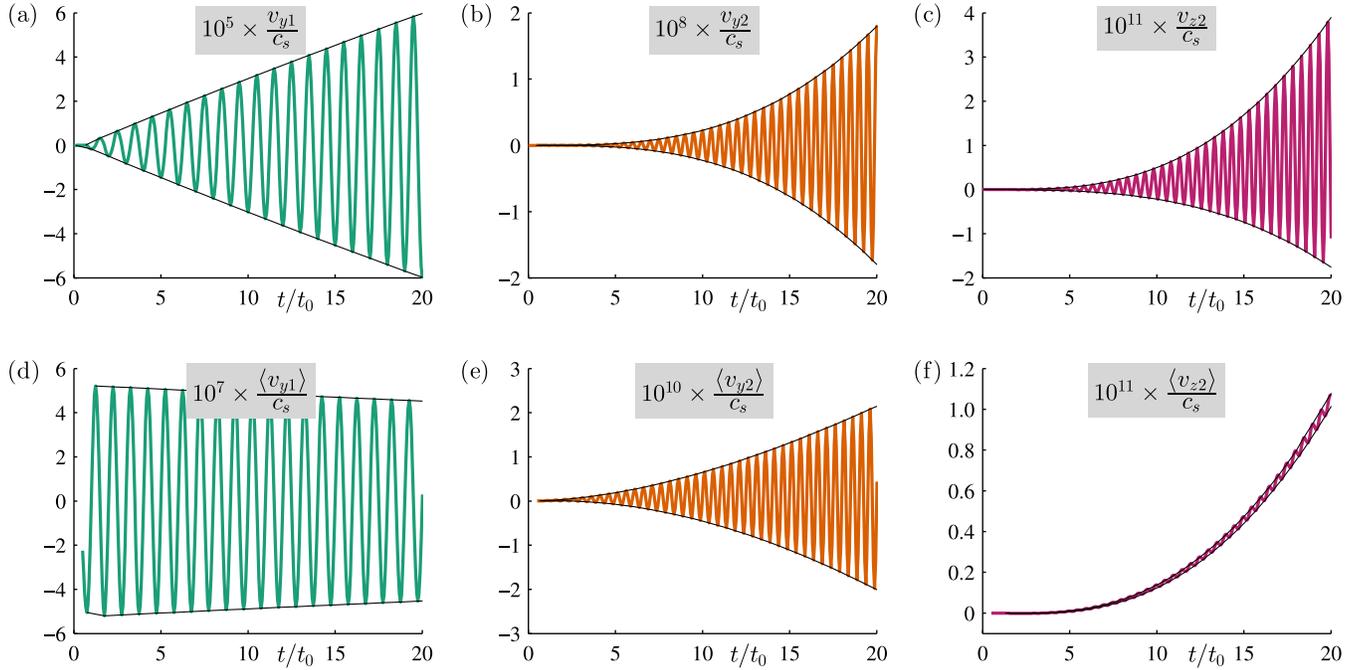}
\caption{\figlab{short} (Color online) Velocity probes for the initial time interval $0<t<20\,\tO$. (a-c) probes for the first- and second-order velocity (a) $v_{y1}(0,0)$, (b) $v_{y2}(w/4,0)$, and (c) $v_{z2}(0,h/4)$. (d-f) Running time-average \eqref{timeint} on an interval one oscillation period wide of the velocity probes in (a-c). The thick lines show the oscillating velocity probes while the thin lines emphasize the envelopes of the oscillations.}
\end{figure*}
\begin{figure*}
\centering
\includegraphics[width=0.99\textwidth]{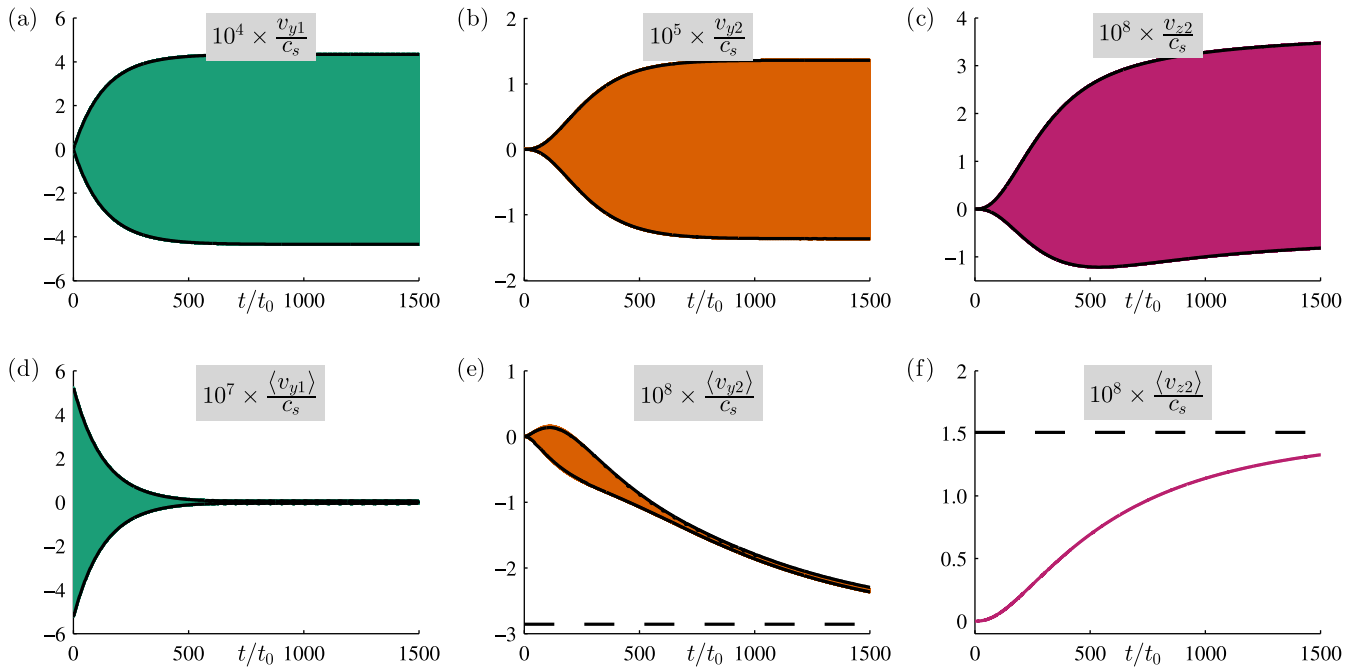}
\caption{\figlab{long} (Color online) The velocity probes from \figref{short}, but now extended to the long time interval $0<t<1500\,\tO$, showing the convergence towards a periodic state. The dashed lines in (e) and (f) indicate the magnitude of the steady time-averaged second-order velocity from the frequency-domain simulation \eqsref{freqdom2ndmass}{freqdom2ndmom}.}
\end{figure*}
%

%
\subsection{Single-pulse scaling analysis}

A striking feature of \figref{long}, is the separation of timescales between the roughly exponential build-up of the acoustic resonance in \figref{long}(a) and of the streaming flow in \figref{long}(f). It appears that the resonance, and hence the acoustic radiation force on a suspended particle, is fully established almost ten times faster than the streaming flow and the resulting drag force on a suspended particle. To investigate this further, we look at the scaling provided by the three timescales relevant for the problem of transient acoustic streaming, all listed in \tabref{timescales}: the oscillation time  $\tO$ of the acoustic wave, the resonance relaxation time $\tauE$ of the acoustic cavity, and the momentum diffusion time $\taunu$ governing the quasi-steady streaming flow.

\begin{table}
\caption{\tablab{timescales}
Characteristic timescales. The values are obtained by using the kinematic viscosity $\nu = \eta/\rhoO = 8.93\times 10^{-7}~\SIm^2/\SIs$ (\tabref{parameter_values}), the Q-factor  $Q = 416$ (\figref{resonance}), and the channel height
$h = 160~\SImum$ (\figref{chipmesh}).
}
\begin{tabular}{ L{3.0cm} c  r@{ $\times$ }l@{ s}  r@{ $\approx$ } r@{ $\tO$}}
\hline \hline
Timescale & Expression & \multicolumn{2}{c}{Value} & \multicolumn{2}{c}{} \\ \hline
Oscillation time & $\tO$ & $5.1$ & $10^{-7}$ & & $1$ \Tstrut\\[1.0mm]
Resonance\newline relaxation time & $\tauE=\frac{Q}{2\pi}t_0$ & $3.4$ & $10^{-5}$ & & $66$ \\[2.0mm]
Momentum \newline diffusion time & $\taunu=\frac{1}{2\nu}\big(\frac{h}{8}\big)^2$ & $2.8$ & $10^{-4}$ & & $558$\\[2.5mm]
\hline \hline
\end{tabular}
\end{table}

The momentum diffusion time is  $\taunu = \frac{1}{2\nu}\big(\frac{1}{8}h\big)^2$, where $\nu=\frac{\eta}{\rhoO}$ is the kinematic viscosity, and $\frac18 h$ is approximately half the distance between the top boundary and the center of the streaming flow roll. Inserting the relevant numbers, see \tabref{timescales}, we indeed find that $\tau_E \approx 66~\tO$ is much faster than $\taunu \approx 558\tO$. However, this separation in timescales does not guarantee a suppression of streaming relative to the radiation force. One problem is that the streaming is driven by the shear stresses in the boundary layer, and these stresses builds up much faster given the small thickness of the boundary layer. This we investigate further in the following subsection. Another problem is that the large momentum diffusion time $\taunu$ implies a very slow decay of the streaming flow, once it is established. The latter effect, we study using the following analytical model. Consider a quantity $f$ (streaming velocity or acoustic energy), with a relaxation time $\tau$ and driven by a pulsed source term $P$ of pulse width $\tpw$. The rate of change of $f$ is equivalent to \eqref{Ebalance},
\bsubal
\eqlab{resonatorEq}
\pp_t f &=  P - \frac{1}{\tau} f,\\
P &= \left\{ \begin{array}{cl} \frac{1}{\tau}f_0, &\text{ for }\; 0 < t < \tpw, \\[2mm] 0, & \text{ otherwise}, \end{array} \right.
\esubal
where $\frac{1}{\tau}f_0$ is a constant input power. This simplified analytical model captures the roughly exponential build-up and decay characteristics of our full numerical model, and allows for analytical studies of the time-integral of $f(t)$. For a final time $t > \tpw$ we find
\bal \eqlab{integral}
\int_0^t f(t')\, \dd t' = f_0\tpw - f_0\tau\Big[\ee^{-\frac{1}{\tau}(t-\tpw)}- \ee^{-\frac{1}{\tau}t}\Big].
\eal
From this we see that when $t \gg \tau + \tpw$  the time-integral of $f(t)$ is approximately $f_0\tpw$ and not dependent on the relaxation time $\tau$. Consequently, if both the acoustic energy and the acoustic streaming can be described by exponential behavior with the respective relaxation times $\tau_E$ and $\taunu$, the ratio of their time-integrated effects is the same whether the system is driven by a constant actuation towards their steady time-periodic state or by a pulsed actuation with pulse width $\tpw$. This simplified analytical model indicates that there is little hope of decreasing acoustic streaming relative to the acoustic radiation force by applying pulsed actuation, in spite of the order of magnitude difference between the relaxation times for the acoustic energy and the streaming.

%
\subsection{Single-pulse numerical analysis}

We investigate the features of pulsed actuation more detailed in the following by numerical analysis. In \figref{pulse} is shown the temporal evolution of the total acoustic energy $\avr{\Eac}$ and the magnitude of the acoustic streaming flow $\avr{\vstr}$ for the three cases: (i) the build-up towards the periodic state, (ii) a single long actuation pulse, and (iii) a single short actuation pulse. The magnitude of the acoustic streaming is measured by the unsteady time-averaged velocity probe
\beq{vstr}
\avr{\vstr} =\avr{\vzII(0,\frac{1}{4}h)},
\eeq
and the unsteady energy and streaming probes obtained from the time-domain simulation are normalized by their corresponding steady time-averaged values from the frequency-domain simulation.

We introduce the streaming ratio $\chi$ to measure the influence of streaming-induced drag on suspended particles relative to the influence of the acoustic radiation force for the unsteady time-domain solution, in comparison to the  periodic frequency-domain solution. To calculate the relative displacement $\Delta s$ of particles due to each of the two forces, respectively, we compare their time integrals. Since the radiation force scales with the acoustic energy density, we define the streaming ratio $\chi(t)$ as
\bal \eqlab{chi}
\chi(t)= \frac{\displaystyle\int_{0}^t \frac{\avr{v_\mathrm{str}(t')}}{\avr{v_\mathrm{str}^\mathrm{fd}(\infty)}}\ \mathrm{d}t'}
{\displaystyle\int_{0}^t \frac{\avr{\Eac(t')}}{\avr{E_\mathrm{ac}^\mathrm{fd}(\infty)}}\ \mathrm{d}t'}
\sim
\frac{\displaystyle \hspace{1mm}\frac{\Delta s^{\phantom{fd}}_\mathrm{str}}{\Delta s^\mathrm{fd}_\mathrm{str}}\hspace{1mm}}
{\displaystyle \hspace{1mm}\frac{\Delta s^{\phantom{fd}}_\mathrm{rad}}{\Delta s^\mathrm{fd}_\mathrm{rad}}\hspace{1mm}},
\eal
where $\Delta s_\mathrm{str}$ and $\Delta s_\mathrm{rad}$ are the total particle displacements in the time from 0 to $t$ due to the streaming-induced drag force and the acoustic radiation force, respectively. In the periodic state $\chi=1$, and to obtain radiation force-dominated motion of smaller particles, we need to achieve a smaller value of $\chi$.
Obtaining a value of $\chi=0.8$ at time $t^{{}}_\mathrm{end}$, implies that the ratio of the relative displacement due to the streaming-induced drag force and the radiation force for the time interval $0<t<t^{{}}_\mathrm{end}$ is 20\% lower than in the periodic state, corresponding to a 20\% reduction of the critical particle size for acoustophoretic focusing, defined in Ref. \cite{Muller2012}, assuming the particles can be focused during the time interval ${0<t<t^{{}}_\mathrm{end}}$.
\begin{figure}
\centering
\includegraphics[width=0.99\columnwidth]{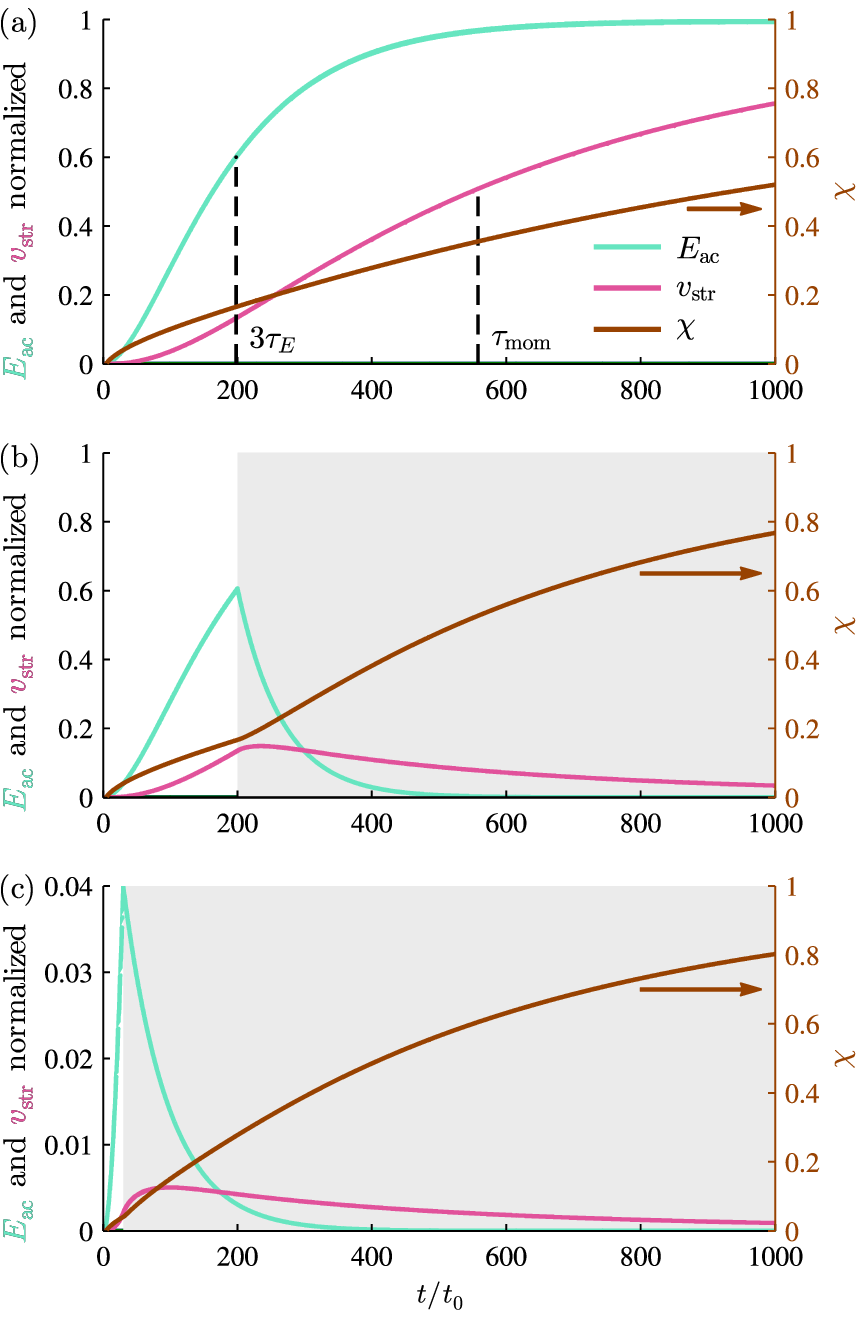}
\caption{\figlab{pulse} (Color online) Acoustic energy $\avr{\Eac(t)}/\avr{E_\mathrm{ac}^\mathrm{fd}(\infty)}$ \eqref{Eacboth} (light green), streaming velocity $\avr{\vstr(t)}/\avr{v_\mathrm{str}^\mathrm{fd}(\infty)}$ \eqref{vstr} (medium purple), and streaming ratio $\chi(t)$ \eqref{chi} (dark brown, right ordinate axis). The gray background indicates the time intervals where the actuation is turned off. (a) Constant actuation for $0<t<3000\,\tO$. (b) Actuation on for $0<t<200\,\tO$ followed by no actuation for $200\,\tO<t<1000\,\tO$. (c) Actuation on for $0<t<30\,\tO$ followed by no actuation for $30\,\tO<t<1000\,\tO$.}
\end{figure}

\begin{figure}
\centering
\includegraphics[width=0.99\columnwidth]{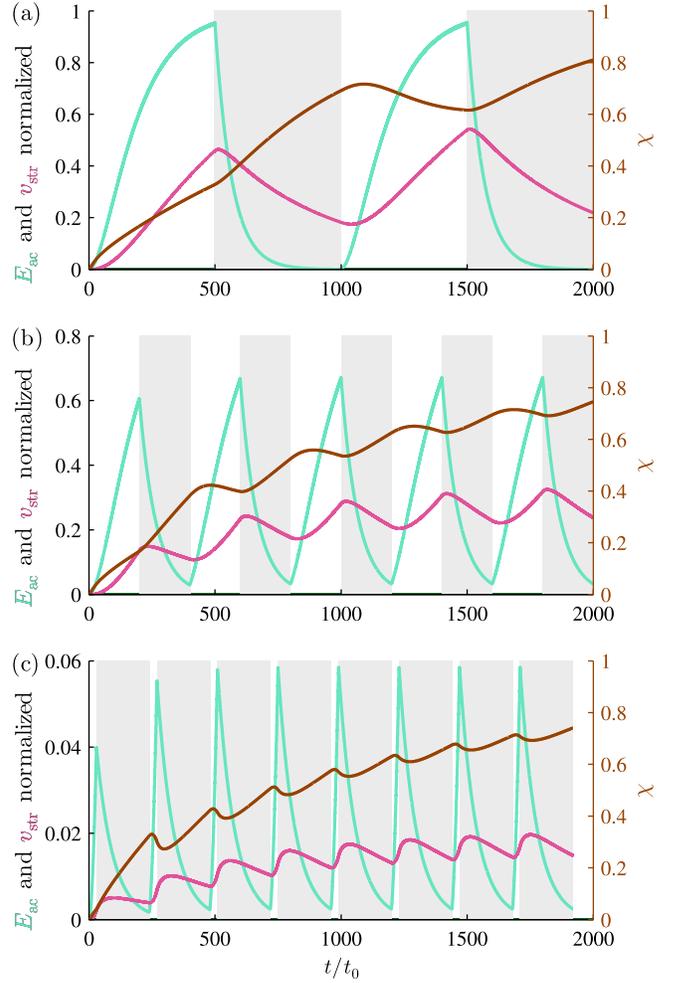}
\caption{\figlab{pulsed} (Color online)  The same probes as in \figref{pulse} but for the following pulsed actuation schemes: (a) actuation is on for $500\,\tO$ followed by no actuation for $500\,\tO$ repeatedly, (b) actuation is on for $200\,\tO$ followed by no actuation for $200\,\tO$ repeatedly, and (c) actuation is on for $30\,\tO$ followed by no actuation for $210\,\tO$ repeatedly.}
\end{figure}

Figure \ref{fig:pulse}(a) shows $\avr{\Eac}$, $\avr{\vstr}$ and $\chi$ during the build-up towards the periodic state. $\chi$ approaches unity slower than $\avr{\vstr}$ because $\chi$ is an integration of the streaming and radiation contributions, whereas $\vstr$ probes the instantaneous magnitude of the streaming flow.
Figure \ref{fig:pulse}(b) and \ref{fig:pulse}(c) show $\avr{\Eac}$, $\avr{\vstr}$, and $\chi$ when the actuation is turned off at $t=200\,\tO$ and $t=30\,\tO$, respectively. When the actuation is turned off, $\avr{\Eac}$ decays faster than $\avr{\vstr}$ and thus $\chi$ begins to increase more rapidly, reaching $\chi=0.8$ around $t=1000\,\tO$ in both cases. From the results shown in \figref{pulse} it does not seem advantages to turn off the actuation, as this only causes $\chi$ to increase faster than for constant actuation. Figure \ref{fig:pulse}(c) further shows that when the actuation is turned off, $\avr{\Eac}$ immediately begins to decay, whereas $\avr{\vstr}$ continues to increase for some time, due to the present acoustic energy in the system that still provides a driving force for the streaming flow.

%
\subsection{Multi-pulse numerical analysis}
From the single pulse results shown in \figref{pulse} there is no indication of any optimum for the pulse duration or repetition period, and in general it provides little hope that pulsed actuation should lead to lower values of $\chi$. Figure \ref{fig:pulsed} shows $\avr{\Eac}$, $\avr{\vstr}$, and $\chi$ for three pulsed schemes with pulse duration $500\,\tO$, $200\,\tO$, and $30\,\tO$ and pause duration $500\,\tO$, $200\,\tO$, and $210\,\tO$, respectively. For all three pulsed schemes, $\chi$ increases faster than for the constant actuation \figref{pulse}(a), thus not indicating any increased suppression of the streaming.

\section{Discussion}
\seclab{discussion}
Solving numerically the time-dependent problem of the acoustic cavity and the build-up of acoustic streaming, presents new challenges, which are not present in the purely periodic problem. Firstly, the numerical convergence analysis now involves both the spatial and temporal resolutions. This we addressed in a sequential process by first analyzing the spatial mesh with the periodic frequency-domain solution, and thereafter doing a thorough convergence analysis with respect to the temporal resolution.
Secondly, the convergence of the transient solution towards the periodic state was poor for actuation frequencies away from the resonance frequency of the system. This makes off-resonance simulation computationally costly, as it requires a better temporal resolution, and it complicates comparison of simulations at resonance with simulations off resonance.
Thirdly, small numerical errors accumulate during the hundred thousand time steps taken during a simulation from a quiescent state to a purely periodic state. These errors need to be suppressed by the numerical time-domain solver, which in the generalized-alpha solver is done through the alpha parameter. Simulation with higher temporal resolution required lower values of the alpha parameter to have more suppression of accumulated numerical errors.

The model system used in this study is a simplification of an actual device. The vibration of only the side walls, and not the top and bottom walls, stands in contrast to the physical system, in which the whole device is vibrating in a non-trivial way, difficult to predict, and only the overall amplitude and the frequency of the actuation is controlled experimentally. Furthermore, our model only treats the two-dimensional cross section of a long straight channel, whereas experimental studies have shown that there are dynamics along the length of the channel \cite{Augustsson2011}. Nevertheless, successful comparison, both qualitatively and quantitatively, have been reported between the prediction of this simplified numerical model and experimental measurements of Rayleigh streaming in the cross sectional plane of a microchannel \cite{Muller2013}, which makes it reasonable to assume that the time-dependent simulations also provide reliable predictions.

It is also important to stress that our model only describes the fluid and not the motion of the suspended particles. Integrating the forces acting on the particles becomes vastly more demanding when the streaming flow is unsteady, because the drag forces from the oscillating velocity components $\vvvI$ and $\vvvIIw$ do average out, as they do in the case of a purely time-periodic state. To include this contribution in the particle tracking scheme, the forces on the particles need to be integrated with a time step of a fraction of the oscillation period, which makes the solution of particle trajectories over several seconds a very demanding task using brute-force integration of the equations of motion.

Our analysis of the pulsed actuation schemes showed that the slow decay of the streaming flow makes pulsation inefficient in reducing the streaming-induced drag force compared to the radiation force. Such a reduction may, however, be obtained by a rapid switching between different resonances each resulting in similar radiation forces but different spatial streaming patterns which on averages cancel each other out, thus fighting streaming with streaming. An idea along these lines was presented by Ohlin \etal\ Ref.~\cite{Ohlin2011}, who used frequency sweeping to diminish the streaming flows in liquid-filled wells in a multi-well plate for cell analysis. However, the prediction of particle trajectories under such multi-resonance conditions requires an extensive study as described above.

Experimentally, the use of pulsed actuation to decrease streaming flow has been reported by Hoyos \etal\ Ref. \cite{Hoyos2013}. However, this study is not directly comparable to our analysis, as we treat the build-up of Rayleigh streaming perpendicular to the pressure nodal plane, whereas Hoyos \etal \ study the streaming flow in this plane. Such in-nodal-plane streaming flows have been studied numerically by Lei \etal\ \cite{Lei2013,Lei2014}, though only with steady actuation. The contradicting results of our theoretical study and the experimental study of Hoyos \etal\ may thus rely on the differences of the phenomena studied.

\section{Conclusion}
\seclab{conclusion}
In this work, we have presented a model for the transient acoustic fields and the unsteady time-averaged second-order velocity field in the transverse cross-sectional plane of a long straight microchannel. The model is based on the usual perturbation approach for low acoustic field amplitudes, and we have solved both first- and second-order equations in the time domain for the unsteady transient case as well as in the frequency domain for the purely periodic case. This enabled us to characterize the build-up of the oscillating acoustic fields and the unsteady streaming flow.

Our analysis showed that the build-up of acoustic energy in the channel follows the analytical prediction obtained for a single damped harmonic oscillator with sinusoidal forcing, and that a quasi-steady velocity component is established already within the first few oscillations and increases in magnitude as the acoustic energy builds up. We have also found that for a resonance with quality factor $Q$, the amplitude of the oscillatory second-order velocity component is a factor of $Q$ larger than what is expected from dimensional analysis, which results in a more restrictive criterion for the validity of the perturbation expansion, compared to the usual one based on the first-order perturbation expansion.

Furthermore, contrary to a simple scaling analysis of the time scales involved in the fast build-up of radiation forces and slow build-up of drag-induced streaming forces, we have found that pulsating oscillatory boundary actuation does not reduce the time-integrated streaming-induced drag force relative to the time-integrated radiation force. As a result, pulsating actuation does not prevent streaming flows perpendicular to the pressure nodal plane from destroying the ability to focus small particles by acoustophoresis.

\begin{acknowledgments}
This work was supported by the Danish Council for Independent Research, Technology, and Production Sciences (grant no. 11-107021).

\end{acknowledgments}

\appendix
\section{Amplitude of the second-order oscillatory velocity field}
%
Extending to second order the one-dimensional example given in Ref.~\cite{Bruus2012}, we derive in this appendix the order of magnitude of the second-order oscillatory component $\vIIw$, which was stated in \eqref{decomAmp}.

Like $\avr{g^\notop_2}$ denotes time-averaging over one oscillation period, \eqref{timeint}, and in the periodic state equals the zero-order temporal Fourier component of the field, then $g^{2\omega}_2(\rrr)$ denotes the complex amplitude of the oscillatory second-order mode and is given by the second-order Fourier component
\bal
g^{2\omega}_2(\rrr) = \frac{1}{T}\int_{t-T/2}^{t+T/2}\, g^\notop_2(\rrr,t')\ee^{-i2\omega t'}\, \dd t'.
\eal
By using the general formula for the real part of any complex number $Z$, ${\re[Z] = \tfrac{1}{2}(Z+Z^*)}$, the product $A(\rrr,t)B(\rrr,t)$ of two oscillating fields ${A(\rrr,t) = \re\big[A\eiwt\big]}$ and ${B(\rrr,t) = \re\big[B\eiwt\big]}$ can be decomposed into a steady component and an oscillatory component
\bal \eqlab{AB}
A(t)B(t) &= \tfrac{1}{2}\Big(A\eiwt+A^*\epiwt\Big)\tfrac{1}{2}\Big(B\eiwt+B^*\epiwt\Big) \nn\\
				 &=\tfrac{1}{2}\re\Big[A^*B\Big] + \tfrac{1}{2}\re\Big[AB\eiIIwt \Big],
\eal
from which we introduce the following notation
\bal
\avr{AB} &\equiv \tfrac{1}{2}\re\Big[A^*B\Big]\eqlab{avr},\\
\big(AB\big)^{2\omega} &\equiv \tfrac{1}{2}AB\eqlab{IIw},
\eal
where $A$ and $B$ could be any first-order fields.

The governing equations for the oscillatory second-order component $\vvvIIw$ can be derived from \eqsref{massEq2}{momentumEq2} and in the one-dimensional problem treated in Ref. \cite{Bruus2012}, where the top and bottom walls are not taken into account, they become
\bsub\eqlab{plates_govIIw1}
\bal
-\ii 2\omega \kaps\pIIw &= -\pp_y \vIIw - \kaps\IIw{\vI\pp_y \pI}\eqlab{plates_continuity2_2w}\\
-\ii 2\omega\rhoO\vIIw  &= -\pp_y \pIIw + \left(\tfrac{4}{3}\eta+\etab\right)\ppsqr_y\vIIw \nn\\
& \hspace{4mm}-\IIw{\rhoI(-\ii\omega \vI)} - \rhoO\IIw{\vI\pp_y\vI}.\eqlab{plates_NS2_2w}
\eal
\esub
Applying the $2\omega$-rule of \eqref{IIw} and the mass continuity \eqref{massEq1}, the two last terms of \eqref{plates_NS2_2w} cancel. Inserting \eqref{plates_continuity2_2w} into \eqref{plates_NS2_2w}, the governing equation for $\vIIw$ becomes
\bal\eqlab{plates_govIIw2}
4\kOsqr\vIIw + (1-\ii4\gam)\ppsqr_y \vIIw + \frac{1}{2}\kaps\pp_y(\vI\pp_y\pI) = 0,
\eal
where $\gam$ is the non-dimensional bulk damping coefficient given by ${\gam = \frac{\omega\eta}{2\rhoO\cssqr}\Big(\frac{4}{3}+\frac{\etab}{\eta}\Big)}$, and ${\kO=\frac{\omega}{\cs}}$ is the wavenumber. For the fundamental half-wave resonance, the spatial dependence of the source term $\pp_y(\vI\pp_y\pI)$ is $\sin(2\kO y)$, and the guess for the inhomogeneous solution to \eqref{plates_govIIw2} thus becomes
\beq{plates_vIIw_inhom}
\vIIwinhom = C\sin(2\kO y).
\eeq
Inserting the inhomogeneous solution \eqref{plates_vIIw_inhom} into the governing equation \refeq{plates_govIIw2}, we note that the first term cancels with the ``1'' in the parentheses of the second term, and the order of magnitude of the inhomogeneous solution thus becomes
\bal\eqlab{vIIw_inhom2}
|\vIIw| = C\sim \frac{1}{\gam}\kaps|\vI||\pI| \sim \frac{1}{\gam^3}\frac{\vbcsqr}{\cs} \sim Q^3\frac{\vbcsqr}{\cs},
\eal
which is the result stated in \eqref{decomAmp}.
%
%
%

\begin{thebibliography}{32}%
\makeatletter
\providecommand \@ifxundefined [1]{%
 \@ifx{#1\undefined}
}%
\providecommand \@ifnum [1]{%
 \ifnum #1\expandafter \@firstoftwo
 \else \expandafter \@secondoftwo
 \fi
}%
\providecommand \@ifx [1]{%
 \ifx #1\expandafter \@firstoftwo
 \else \expandafter \@secondoftwo
 \fi
}%
\providecommand \natexlab [1]{#1}%
\providecommand \enquote  [1]{``#1''}%
\providecommand \bibnamefont  [1]{#1}%
\providecommand \bibfnamefont [1]{#1}%
\providecommand \citenamefont [1]{#1}%
\providecommand \href@noop [0]{\@secondoftwo}%
\providecommand \href [0]{\begingroup \@sanitize@url \@href}%
\providecommand \@href[1]{\@@startlink{#1}\@@href}%
\providecommand \@@href[1]{\endgroup#1\@@endlink}%
\providecommand \@sanitize@url [0]{\catcode `\\12\catcode `\$12\catcode
  `\&12\catcode `\#12\catcode `\^12\catcode `\_12\catcode `\%12\relax}%
\providecommand \@@startlink[1]{}%
\providecommand \@@endlink[0]{}%
\providecommand \url  [0]{\begingroup\@sanitize@url \@url }%
\providecommand \@url [1]{\endgroup\@href {#1}{\urlprefix }}%
\providecommand \urlprefix  [0]{URL }%
\providecommand \Eprint [0]{\href }%
\providecommand \doibase [0]{http://dx.doi.org/}%
\providecommand \selectlanguage [0]{\@gobble}%
\providecommand \bibinfo  [0]{\@secondoftwo}%
\providecommand \bibfield  [0]{\@secondoftwo}%
\providecommand \translation [1]{[#1]}%
\providecommand \BibitemOpen [0]{}%
\providecommand \bibitemStop [0]{}%
\providecommand \bibitemNoStop [0]{.\EOS\space}%
\providecommand \EOS [0]{\spacefactor3000\relax}%
\providecommand \BibitemShut  [1]{\csname bibitem#1\endcsname}%
\let\auto@bib@innerbib\@empty
\bibitem [{\citenamefont {Bruus}\ \emph {et~al.}(2011)\citenamefont {Bruus},
  \citenamefont {Dual}, \citenamefont {Hawkes}, \citenamefont {Hill},
  \citenamefont {Laurell}, \citenamefont {Nilsson}, \citenamefont {Radel},
  \citenamefont {Sadhal},\ and\ \citenamefont {Wiklund}}]{Bruus2011c}%
  \BibitemOpen
  \bibfield  {author} {\bibinfo {author} {\bibfnamefont {H.}~\bibnamefont
  {Bruus}}, \bibinfo {author} {\bibfnamefont {J.}~\bibnamefont {Dual}},
  \bibinfo {author} {\bibfnamefont {J.}~\bibnamefont {Hawkes}}, \bibinfo
  {author} {\bibfnamefont {M.}~\bibnamefont {Hill}}, \bibinfo {author}
  {\bibfnamefont {T.}~\bibnamefont {Laurell}}, \bibinfo {author} {\bibfnamefont
  {J.}~\bibnamefont {Nilsson}}, \bibinfo {author} {\bibfnamefont
  {S.}~\bibnamefont {Radel}}, \bibinfo {author} {\bibfnamefont
  {S.}~\bibnamefont {Sadhal}}, \ and\ \bibinfo {author} {\bibfnamefont
  {M.}~\bibnamefont {Wiklund}},\ }\href {\doibase 10.1039/c1lc90058g}
  {\bibfield  {journal} {\bibinfo  {journal} {Lab Chip}\ }\textbf {\bibinfo
  {volume} {11}},\ \bibinfo {pages} {3579} (\bibinfo {year}
  {2011})}\BibitemShut {NoStop}%
\bibitem [{\citenamefont {Muller}\ \emph {et~al.}(2012)\citenamefont {Muller},
  \citenamefont {Barnkob}, \citenamefont {Jensen},\ and\ \citenamefont
  {Bruus}}]{Muller2012}%
  \BibitemOpen
  \bibfield  {author} {\bibinfo {author} {\bibfnamefont {P.~B.}\ \bibnamefont
  {Muller}}, \bibinfo {author} {\bibfnamefont {R.}~\bibnamefont {Barnkob}},
  \bibinfo {author} {\bibfnamefont {M.~J.~H.}\ \bibnamefont {Jensen}}, \ and\
  \bibinfo {author} {\bibfnamefont {H.}~\bibnamefont {Bruus}},\ }\href
  {\doibase 10.1039/C2LC40612H} {\bibfield  {journal} {\bibinfo  {journal} {Lab
  Chip}\ }\textbf {\bibinfo {volume} {12}},\ \bibinfo {pages} {4617} (\bibinfo
  {year} {2012})}\BibitemShut {NoStop}%
\bibitem [{\citenamefont {Barnkob}\ \emph {et~al.}(2012)\citenamefont
  {Barnkob}, \citenamefont {Augustsson}, \citenamefont {Laurell},\ and\
  \citenamefont {Bruus}}]{Barnkob2012a}%
  \BibitemOpen
  \bibfield  {author} {\bibinfo {author} {\bibfnamefont {R.}~\bibnamefont
  {Barnkob}}, \bibinfo {author} {\bibfnamefont {P.}~\bibnamefont {Augustsson}},
  \bibinfo {author} {\bibfnamefont {T.}~\bibnamefont {Laurell}}, \ and\
  \bibinfo {author} {\bibfnamefont {H.}~\bibnamefont {Bruus}},\ }\href
  {\doibase 10.1103/PhysRevE.86.056307} {\bibfield  {journal} {\bibinfo
  {journal} {Phys Rev E}\ }\textbf {\bibinfo {volume} {86}},\ \bibinfo {pages}
  {056307} (\bibinfo {year} {2012})}\BibitemShut {NoStop}%
\bibitem [{\citenamefont {Rayleigh}(1884)}]{LordRayleigh1884}%
  \BibitemOpen
  \bibfield  {author} {\bibinfo {author} {\bibfnamefont {L.}~\bibnamefont
  {Rayleigh}},\ }\href@noop {} {\bibfield  {journal} {\bibinfo  {journal}
  {Philos Trans R Soc London}\ }\textbf {\bibinfo {volume} {175}},\ \bibinfo
  {pages} {1} (\bibinfo {year} {1884})}\BibitemShut {NoStop}%
\bibitem [{\citenamefont {Schlichting}(1932)}]{Schlichting1932}%
  \BibitemOpen
  \bibfield  {author} {\bibinfo {author} {\bibfnamefont {H.}~\bibnamefont
  {Schlichting}},\ }\href@noop {} {\bibfield  {journal} {\bibinfo  {journal}
  {Phys Z}\ }\textbf {\bibinfo {volume} {33}},\ \bibinfo {pages} {327}
  (\bibinfo {year} {1932})}\BibitemShut {NoStop}%
\bibitem [{\citenamefont {Nyborg}(1958)}]{Nyborg1958}%
  \BibitemOpen
  \bibfield  {author} {\bibinfo {author} {\bibfnamefont {W.~L.}\ \bibnamefont
  {Nyborg}},\ }\href {\doibase {10.1121/1.1909587}} {\bibfield  {journal}
  {\bibinfo  {journal} {J Acoust Soc Am}\ }\textbf {\bibinfo {volume} {30}},\
  \bibinfo {pages} {329} (\bibinfo {year} {1958})}\BibitemShut {NoStop}%
\bibitem [{\citenamefont {Hamilton}\ \emph
  {et~al.}(2003{\natexlab{a}})\citenamefont {Hamilton}, \citenamefont
  {Ilinskii},\ and\ \citenamefont {Zabolotskaya}}]{Hamilton2003}%
  \BibitemOpen
  \bibfield  {author} {\bibinfo {author} {\bibfnamefont {M.}~\bibnamefont
  {Hamilton}}, \bibinfo {author} {\bibfnamefont {Y.}~\bibnamefont {Ilinskii}},
  \ and\ \bibinfo {author} {\bibfnamefont {E.}~\bibnamefont {Zabolotskaya}},\
  }\href {\doibase 10.1121/1.1528928} {\bibfield  {journal} {\bibinfo
  {journal} {J Acoust Soc Am}\ }\textbf {\bibinfo {volume} {113}},\ \bibinfo
  {pages} {153} (\bibinfo {year} {2003}{\natexlab{a}})}\BibitemShut {NoStop}%
\bibitem [{\citenamefont {Hamilton}\ \emph
  {et~al.}(2003{\natexlab{b}})\citenamefont {Hamilton}, \citenamefont
  {Ilinskii},\ and\ \citenamefont {Zabolotskaya}}]{Hamilton2003a}%
  \BibitemOpen
  \bibfield  {author} {\bibinfo {author} {\bibfnamefont {M.}~\bibnamefont
  {Hamilton}}, \bibinfo {author} {\bibfnamefont {Y.}~\bibnamefont {Ilinskii}},
  \ and\ \bibinfo {author} {\bibfnamefont {E.}~\bibnamefont {Zabolotskaya}},\
  }\href {\doibase {10.1121/1.1618752]}} {\bibfield  {journal} {\bibinfo
  {journal} {J Acoust Soc Am}\ }\textbf {\bibinfo {volume} {114}},\ \bibinfo
  {pages} {3092} (\bibinfo {year} {2003}{\natexlab{b}})}\BibitemShut {NoStop}%
\bibitem [{\citenamefont {Rednikov}\ and\ \citenamefont
  {Sadhal}(2011)}]{Rednikov2011}%
  \BibitemOpen
  \bibfield  {author} {\bibinfo {author} {\bibfnamefont {A.~Y.}\ \bibnamefont
  {Rednikov}}\ and\ \bibinfo {author} {\bibfnamefont {S.~S.}\ \bibnamefont
  {Sadhal}},\ }\href {\doibase 10.1017/S0022112010004532} {\bibfield  {journal}
  {\bibinfo  {journal} {J Fluid Mech}\ }\textbf {\bibinfo {volume} {667}},\
  \bibinfo {pages} {426} (\bibinfo {year} {2011})}\BibitemShut {NoStop}%
\bibitem [{\citenamefont {Muller}\ \emph {et~al.}(2013)\citenamefont {Muller},
  \citenamefont {Rossi}, \citenamefont {Marin}, \citenamefont {Barnkob},
  \citenamefont {Augustsson}, \citenamefont {Laurell}, \citenamefont
  {K\"{a}hler},\ and\ \citenamefont {Bruus}}]{Muller2013}%
  \BibitemOpen
  \bibfield  {author} {\bibinfo {author} {\bibfnamefont {P.~B.}\ \bibnamefont
  {Muller}}, \bibinfo {author} {\bibfnamefont {M.}~\bibnamefont {Rossi}},
  \bibinfo {author} {\bibfnamefont {A.~G.}\ \bibnamefont {Marin}}, \bibinfo
  {author} {\bibfnamefont {R.}~\bibnamefont {Barnkob}}, \bibinfo {author}
  {\bibfnamefont {P.}~\bibnamefont {Augustsson}}, \bibinfo {author}
  {\bibfnamefont {T.}~\bibnamefont {Laurell}}, \bibinfo {author} {\bibfnamefont
  {C.~J.}\ \bibnamefont {K\"{a}hler}}, \ and\ \bibinfo {author} {\bibfnamefont
  {H.}~\bibnamefont {Bruus}},\ }\href {\doibase {10.1103/PhysRevE.88.023006}}
  {\bibfield  {journal} {\bibinfo  {journal} {Phys Rev E}\ }\textbf {\bibinfo
  {volume} {88}},\ \bibinfo {pages} {023006} (\bibinfo {year}
  {2013})}\BibitemShut {NoStop}%
\bibitem [{\citenamefont {Muller}\ and\ \citenamefont
  {Bruus}(2014)}]{Muller2014}%
  \BibitemOpen
  \bibfield  {author} {\bibinfo {author} {\bibfnamefont {P.~B.}\ \bibnamefont
  {Muller}}\ and\ \bibinfo {author} {\bibfnamefont {H.}~\bibnamefont {Bruus}},\
  }\href {\doibase 10.1103/PhysRevE.90.043016} {\bibfield  {journal} {\bibinfo
  {journal} {Phys Rev E}\ }\textbf {\bibinfo {volume} {90}},\ \bibinfo {pages}
  {043016} (\bibinfo {year} {2014})}\BibitemShut {NoStop}%
\bibitem [{\citenamefont {Lei}\ \emph {et~al.}(2013)\citenamefont {Lei},
  \citenamefont {Glynne-Jones},\ and\ \citenamefont {Hill}}]{Lei2013}%
  \BibitemOpen
  \bibfield  {author} {\bibinfo {author} {\bibfnamefont {J.}~\bibnamefont
  {Lei}}, \bibinfo {author} {\bibfnamefont {P.}~\bibnamefont {Glynne-Jones}}, \
  and\ \bibinfo {author} {\bibfnamefont {M.}~\bibnamefont {Hill}},\ }\href
  {\doibase {10.1039/c3lc00010a}} {\bibfield  {journal} {\bibinfo  {journal}
  {Lab Chip}\ }\textbf {\bibinfo {volume} {13}},\ \bibinfo {pages} {2133}
  (\bibinfo {year} {2013})}\BibitemShut {NoStop}%
\bibitem [{\citenamefont {Lei}\ \emph {et~al.}(2014)\citenamefont {Lei},
  \citenamefont {Hill},\ and\ \citenamefont {Glynne-Jones}}]{Lei2014}%
  \BibitemOpen
  \bibfield  {author} {\bibinfo {author} {\bibfnamefont {J.}~\bibnamefont
  {Lei}}, \bibinfo {author} {\bibfnamefont {M.}~\bibnamefont {Hill}}, \ and\
  \bibinfo {author} {\bibfnamefont {P.}~\bibnamefont {Glynne-Jones}},\ }\href
  {\doibase {10.1039/c3lc50985k}} {\bibfield  {journal} {\bibinfo  {journal}
  {Lab Chip}\ }\textbf {\bibinfo {volume} {14}},\ \bibinfo {pages} {532}
  (\bibinfo {year} {2014})}\BibitemShut {NoStop}%
\bibitem [{\citenamefont {Nyborg}(1953)}]{Nyborg1953a}%
  \BibitemOpen
  \bibfield  {author} {\bibinfo {author} {\bibfnamefont {W.~L.}\ \bibnamefont
  {Nyborg}},\ }\href {\doibase {10.1121/1.1907010}} {\bibfield  {journal}
  {\bibinfo  {journal} {J Acoust Soc Am}\ }\textbf {\bibinfo {volume} {25}},\
  \bibinfo {pages} {68} (\bibinfo {year} {1953})}\BibitemShut {NoStop}%
\bibitem [{\citenamefont {Lee}\ and\ \citenamefont {Wang}(1989)}]{Lee1989}%
  \BibitemOpen
  \bibfield  {author} {\bibinfo {author} {\bibfnamefont {C.}~\bibnamefont
  {Lee}}\ and\ \bibinfo {author} {\bibfnamefont {T.}~\bibnamefont {Wang}},\
  }\href {\doibase {10.1121/1.397491}} {\bibfield  {journal} {\bibinfo
  {journal} {J Acoust Soc Am}\ }\textbf {\bibinfo {volume} {85}},\ \bibinfo
  {pages} {1081} (\bibinfo {year} {1989})}\BibitemShut {NoStop}%
\bibitem [{\citenamefont {Hahn}\ and\ \citenamefont {Dual}(2015)}]{Hahn2015}%
  \BibitemOpen
  \bibfield  {author} {\bibinfo {author} {\bibfnamefont {P.}~\bibnamefont
  {Hahn}}\ and\ \bibinfo {author} {\bibfnamefont {J.}~\bibnamefont {Dual}},\
  }\href {\doibase 10.1063/1.4922986} {\bibfield  {journal} {\bibinfo
  {journal} {Physics of Fluids}\ }\textbf {\bibinfo {volume} {27}},\ \bibinfo
  {pages} {062005} (\bibinfo {year} {2015})}\BibitemShut {NoStop}%
\bibitem [{\citenamefont {Nama}\ \emph {et~al.}(2015)\citenamefont {Nama},
  \citenamefont {Barnkob}, \citenamefont {Mao}, \citenamefont {K\"{a}hler},
  \citenamefont {Costanzo},\ and\ \citenamefont {Huang}}]{Nama2015}%
  \BibitemOpen
  \bibfield  {author} {\bibinfo {author} {\bibfnamefont {N.}~\bibnamefont
  {Nama}}, \bibinfo {author} {\bibfnamefont {R.}~\bibnamefont {Barnkob}},
  \bibinfo {author} {\bibfnamefont {Z.}~\bibnamefont {Mao}}, \bibinfo {author}
  {\bibfnamefont {C.~J.}\ \bibnamefont {K\"{a}hler}}, \bibinfo {author}
  {\bibfnamefont {F.}~\bibnamefont {Costanzo}}, \ and\ \bibinfo {author}
  {\bibfnamefont {T.~J.}\ \bibnamefont {Huang}},\ }\href {\doibase
  10.1039/c5lc00231a} {\bibfield  {journal} {\bibinfo  {journal} {Lab Chip}\
  }\textbf {\bibinfo {volume} {15}},\ \bibinfo {pages} {2700} (\bibinfo {year}
  {2015})}\BibitemShut {NoStop}%
\bibitem [{\citenamefont {Hoyos}\ and\ \citenamefont
  {Castro}(2013)}]{Hoyos2013}%
  \BibitemOpen
  \bibfield  {author} {\bibinfo {author} {\bibfnamefont {M.}~\bibnamefont
  {Hoyos}}\ and\ \bibinfo {author} {\bibfnamefont {A.}~\bibnamefont {Castro}},\
  }\href {\doibase 10.1016/j.ultras.2012.03.015} {\bibfield  {journal}
  {\bibinfo  {journal} {Ultrasonics}\ }\textbf {\bibinfo {volume} {53}},\
  \bibinfo {pages} {70} (\bibinfo {year} {2013})}\BibitemShut {NoStop}%
\bibitem [{\citenamefont {Antfolk}\ \emph {et~al.}(2014)\citenamefont
  {Antfolk}, \citenamefont {Muller}, \citenamefont {Augustsson}, \citenamefont
  {Bruus},\ and\ \citenamefont {Laurell}}]{Antfolk2014}%
  \BibitemOpen
  \bibfield  {author} {\bibinfo {author} {\bibfnamefont {M.}~\bibnamefont
  {Antfolk}}, \bibinfo {author} {\bibfnamefont {P.~B.}\ \bibnamefont {Muller}},
  \bibinfo {author} {\bibfnamefont {P.}~\bibnamefont {Augustsson}}, \bibinfo
  {author} {\bibfnamefont {H.}~\bibnamefont {Bruus}}, \ and\ \bibinfo {author}
  {\bibfnamefont {T.}~\bibnamefont {Laurell}},\ }\href {\doibase DOI:
  10.1039/c4lc00202d} {\bibfield  {journal} {\bibinfo  {journal} {Lab Chip}\
  }\textbf {\bibinfo {volume} {14}},\ \bibinfo {pages} {2791} (\bibinfo {year}
  {2014})}\BibitemShut {NoStop}%
\bibitem [{\citenamefont {Wang}\ and\ \citenamefont {Dual}(2009)}]{Wang2009}%
  \BibitemOpen
  \bibfield  {author} {\bibinfo {author} {\bibfnamefont {J.}~\bibnamefont
  {Wang}}\ and\ \bibinfo {author} {\bibfnamefont {J.}~\bibnamefont {Dual}},\
  }\href@noop {} {\bibfield  {journal} {\bibinfo  {journal} {J. Phys. A: Math.
  Theor.}\ }\textbf {\bibinfo {volume} {42}},\ \bibinfo {pages} {285502}
  (\bibinfo {year} {2009})}\BibitemShut {NoStop}%
\bibitem [{\citenamefont {Pierce}(1991)}]{Pierce1991}%
  \BibitemOpen
  \bibfield  {author} {\bibinfo {author} {\bibfnamefont {A.~D.}\ \bibnamefont
  {Pierce}},\ }\href@noop {} {\emph {\bibinfo {title} {Acoustics}}}\ (\bibinfo
  {publisher} {Acoustical Society of America},\ \bibinfo {address} {Woodbury},\
  \bibinfo {year} {1991})\BibitemShut {NoStop}%
\bibitem [{\citenamefont {Landau}\ and\ \citenamefont
  {Lifshitz}(1980)}]{Landau1980}%
  \BibitemOpen
  \bibfield  {author} {\bibinfo {author} {\bibfnamefont {L.~D.}\ \bibnamefont
  {Landau}}\ and\ \bibinfo {author} {\bibfnamefont {E.~M.}\ \bibnamefont
  {Lifshitz}},\ }\href@noop {} {\emph {\bibinfo {title} {Statistical Physics,
  Part~1}}},\ \bibinfo {edition} {3rd}\ ed.,\ Vol.~\bibinfo {volume} {5}\
  (\bibinfo  {publisher} {Butterworth-Heinemann},\ \bibinfo {address}
  {Oxford},\ \bibinfo {year} {1980})\BibitemShut {NoStop}%
\bibitem [{\citenamefont {Augustsson}\ \emph {et~al.}(2011)\citenamefont
  {Augustsson}, \citenamefont {Barnkob}, \citenamefont {Wereley}, \citenamefont
  {Bruus},\ and\ \citenamefont {Laurell}}]{Augustsson2011}%
  \BibitemOpen
  \bibfield  {author} {\bibinfo {author} {\bibfnamefont {P.}~\bibnamefont
  {Augustsson}}, \bibinfo {author} {\bibfnamefont {R.}~\bibnamefont {Barnkob}},
  \bibinfo {author} {\bibfnamefont {S.~T.}\ \bibnamefont {Wereley}}, \bibinfo
  {author} {\bibfnamefont {H.}~\bibnamefont {Bruus}}, \ and\ \bibinfo {author}
  {\bibfnamefont {T.}~\bibnamefont {Laurell}},\ }\href {\doibase
  10.1039/c1lc20637k} {\bibfield  {journal} {\bibinfo  {journal} {Lab Chip}\
  }\textbf {\bibinfo {volume} {11}},\ \bibinfo {pages} {4152} (\bibinfo {year}
  {2011})}\BibitemShut {NoStop}%
\bibitem [{\citenamefont {{COMSOL Multiphysics 4.4,
  www.comsol.com}}(2013)}]{COMSOL44}%
  \BibitemOpen
  \bibfield  {author} {\bibinfo {author} {\bibnamefont {{COMSOL Multiphysics
  4.4, www.comsol.com}}},\ }\href@noop {} {} (\bibinfo {year}
  {2013})\BibitemShut {NoStop}%
\bibitem [{\citenamefont {Brenner}\ and\ \citenamefont
  {Scott}(2008)}]{Brenner2008}%
  \BibitemOpen
  \bibfield  {author} {\bibinfo {author} {\bibfnamefont {S.~C.}\ \bibnamefont
  {Brenner}}\ and\ \bibinfo {author} {\bibfnamefont {L.~R.}\ \bibnamefont
  {Scott}},\ }\href@noop {} {\emph {\bibinfo {title} {The Mathematical Theory
  of Finite Element Methods}}}\ (\bibinfo  {publisher} {Springer},\ \bibinfo
  {year} {2008})\BibitemShut {NoStop}%
\bibitem [{com()}]{comsolref}%
  \BibitemOpen
  \href@noop {} {\emph {\bibinfo {title} {Comsol Multiphysics Reference Manual
  version 4.4}}}\BibitemShut {NoStop}%
\bibitem [{\citenamefont {{Matlab 2012a,
  www.mathworks.com}}(2012)}]{Matlab2012a}%
  \BibitemOpen
  \bibfield  {author} {\bibinfo {author} {\bibnamefont {{Matlab 2012a,
  www.mathworks.com}}},\ }\href@noop {} {} (\bibinfo {year} {2012})\BibitemShut
  {NoStop}%
\bibitem [{Note1()}]{Note1}%
  \BibitemOpen
  \bibinfo {note} {See Supplemental Material at [URL] for Comsol model files,
  both in a simple version \protect \texttt {Muller2015\protect
  \_TimeDepAcoust\protect \_simple.mph}, and a full version \protect
  \texttt {Muller2015\protect \_TimeDepAcoust\protect \_full.mph}, that
  allows for sectioning in smaller time intervals using the supplied Matlab
  script \protect \texttt {Muller2015\protect \_TimeDepAcoust\protect
  \_full.m} with the functions \protect \texttt {AVGromb16PBM.m} and \protect \texttt {polint.m}}\BibitemShut {NoStop}%
\bibitem [{\citenamefont {{Comsol Multiphysics Model
  Library}}(2013)}]{comsolnonlinearacoustics}%
  \BibitemOpen
  \bibfield  {author} {\bibinfo {author} {\bibnamefont {{Comsol Multiphysics
  Model Library}}},\ }\href
  {https://www.comsol.dk/model/nonlinear-acoustics-8212-modeling-of-the-1d-westervelt-equation-12783}
  {\enquote {\bibinfo {title} {{Nonlinear Acoustics --- Modeling of the 1D
  Westervelt Equation}},}\ } (\bibinfo {year} {2013})\BibitemShut {NoStop}%
\bibitem [{\citenamefont {Bruus}(2012)}]{Bruus2012}%
  \BibitemOpen
  \bibfield  {author} {\bibinfo {author} {\bibfnamefont {H.}~\bibnamefont
  {Bruus}},\ }\href {\doibase 10.1039/C1LC20770A} {\bibfield  {journal}
  {\bibinfo  {journal} {Lab Chip}\ }\textbf {\bibinfo {volume} {12}},\ \bibinfo
  {pages} {20} (\bibinfo {year} {2012})}\BibitemShut {NoStop}%
\bibitem [{\citenamefont {Press}\ \emph {et~al.}(2002)\citenamefont {Press},
  \citenamefont {Teukolsky}, \citenamefont {Vetterling},\ and\ \citenamefont
  {Flannery}}]{Press2002}%
  \BibitemOpen
  \bibfield  {author} {\bibinfo {author} {\bibfnamefont {W.~H.}\ \bibnamefont
  {Press}}, \bibinfo {author} {\bibfnamefont {S.~A.}\ \bibnamefont
  {Teukolsky}}, \bibinfo {author} {\bibfnamefont {W.~T.}\ \bibnamefont
  {Vetterling}}, \ and\ \bibinfo {author} {\bibfnamefont {B.~P.}\ \bibnamefont
  {Flannery}},\ }\href@noop {} {\emph {\bibinfo {title} {Numerical Recipies in
  C - The Art of Scientific Computing, 2nd edition}}}\ (\bibinfo  {publisher}
  {Cambridge University Press},\ \bibinfo {year} {2002})\BibitemShut {NoStop}%
\bibitem [{\citenamefont {Ohlin}\ \emph {et~al.}(2011)\citenamefont {Ohlin},
  \citenamefont {Christakou}, \citenamefont {Frisk}, \citenamefont
  {\"{O}nfelt},\ and\ \citenamefont {Wiklund}}]{Ohlin2011}%
  \BibitemOpen
  \bibfield  {author} {\bibinfo {author} {\bibfnamefont {M.}~\bibnamefont
  {Ohlin}}, \bibinfo {author} {\bibfnamefont {A.}~\bibnamefont {Christakou}},
  \bibinfo {author} {\bibfnamefont {T.}~\bibnamefont {Frisk}}, \bibinfo
  {author} {\bibfnamefont {B.}~\bibnamefont {\"{O}nfelt}}, \ and\ \bibinfo
  {author} {\bibfnamefont {M.}~\bibnamefont {Wiklund}},\ }in\ \href@noop {}
  {\emph {\bibinfo {booktitle} {Proc. 15th MicroTAS, 2 - 6 October 2011,
  Seattle (WA), USA}}},\ \bibinfo {editor} {edited by\ \bibinfo {editor}
  {\bibfnamefont {J.}~\bibnamefont {Landers}}, \bibinfo {editor} {\bibfnamefont
  {A.}~\bibnamefont {Herr}}, \bibinfo {editor} {\bibfnamefont {D.}~\bibnamefont
  {Juncker}}, \bibinfo {editor} {\bibfnamefont {N.}~\bibnamefont {Pamme}}, \
  and\ \bibinfo {editor} {\bibfnamefont {J.}~\bibnamefont {Bienvenue}}}\
  (\bibinfo  {publisher} {CBMS},\ \bibinfo {year} {2011})\ pp.\ \bibinfo
  {pages} {1612--1614}\BibitemShut {NoStop}%
\end{thebibliography}
%
\end{document}